%% file: main.tex
\documentclass[a4paper,UKenglish]{lipics}
 
\usepackage{microtype}

\usepackage{listings}
\usepackage[colorinlistoftodos]{todonotes}
\usepackage{booktabs}

\usepackage{paralist}
\usepackage{array}
\usepackage{collcell}

\usepackage{tikz}
\usetikzlibrary{shapes,arrows,fit,backgrounds,calc,positioning}

\usepackage{dsfont}

\usepackage{mathpartir}
\usepackage{stmaryrd}
\usepackage{amssymb}

\input{macros}

\title{\JSC{}: Higher-Order Contracts for JavaScript%
  \newline\textnormal{Technical Report}
}
\titlerunning{\JSC{}: Higher-Order Contracts for JavaScript}

\author{Matthias Keil}
\author{Peter Thiemann}
\affil{Institute for Computer Science\\University of Freiburg\\Freiburg, Germany\\
  \texttt{\{keilr,thiemann\}@informatik.uni-freiburg.de}
}

\authorrunning{M. Keil and P. Thiemann}

\Copyright{Matthias Keil and Peter Thiemann}



\begin{document}

\maketitle

\input{abstract}
\input{body}


\appendix
\input{appendix}



\bibliography{main}

\end{document}

%% file: macros.tex
\newcommand{\Label}[1]{\textit{#1}}

\newcommand\JSC{\textit{\textsf{Treat{JS}}}}
\newcommand\TJS{\JSC}

\newcommand\syntax\textit

\newcommand\fulfills\vdash

\newcommand\Sem[1]{\llbracket#1\rrbracket}

\newcommand\Rangeof[1]{\llparenthesis#1\rrparenthesis}

\lstdefinelanguage{JavaScript}{
  keywords={attributes, class, classend, do, empty, endif, endwhile, fail,
    function, functionend, if, implements, in, inherit, inout, not, of,
    operations, out, return, set, then, types, while, use, else, switch, case,
  break, default, for, var},
  keywordstyle=\color{black}\bfseries,
  ndkeywords={assert, construct, Contract, Base, Function, Dependent, Object, AObject, With, And, Or, Not, Intersection, Union, Constructor, build, ctor, SFunction, AFunction, Method, AMethod, SMethod, Map, RegExMap, Mapping, Match, SRecursive},
  ndkeywordstyle=\color{black}\bfseries,
  identifierstyle=\color{black},
  sensitive=false,
  comment=[l]{//},
  morecomment = [s]{/*}{*/},
  morecomment = [s][\color{gray}]{/**}{*/},
  commentstyle=\color{gray},
  stringstyle=\color{black}
}
\newcommand{\df}{:=}
\newcommand\bbc{::=}

\newcommand{\eval}{~\Downarrow~} 
\newcommand{\evalc}[1]{~\Downarrow_{#1}~}

\newcommand{\entails}{~\vdash~}
\newcommand{\entailsC}{~\vdash~}
\newcommand{\entailsFA}{~\vdash~}
\newcommand{\entailsPR}{~\vdash~}
\newcommand{\entailsPA}{~\vdash~}
\newcommand{\entailsWP}{~\vdash~}
\newcommand{\entailsOP}{~\vdash~}

\newcommand{\dom}[1]{\textit{dom(}#1\textit{)}}

\newcommand{\lcon}{\lambda_{J}^{\C on}}

\newcommand{\ljNew}{\textit{new}}

\newcommand{\ljTrue}{\textit{true}}
\newcommand{\ljFalse}{\textit{false}}
\newcommand{\ljUndefined}{\textit{undefined}}
\newcommand{\ljNull}{\textit{null}}
\newcommand{\ljNaN}{\textit{NaN}}
\newcommand{\ljEmyStr}{\textit{``''}}

\newcommand{\ljOp}{\textit{op}}

\newcommand{\ljWrap}{\textit{wrap}}

\newcommand{\ljAt}{@}
\newcommand{\ljAtl}[1]{\ljAt^{\cbBlame_{#1}}}
\newcommand{\ljAtb}[1]{\ljAt^{\cbIdent_{#1}}}
\newcommand{\ljAti}[1]{\ljAt^{\cbVar_{#1}}}

\newcommand{\ljDic}{d}
\newcommand{\ljObj}{o}
\newcommand{\ljVal}{v}
\newcommand{\ljValu}{u}
\newcommand{\ljValw}{w}

\newcommand{\ljSVal}{\widehat{\ljVal}}
\newcommand{\ljSValu}{\widehat{\ljValu}}
\newcommand{\ljSValw}{\widehat{\ljValw}}

\newcommand{\ljConst}{c}
\newcommand{\ljVar}{x}
\newcommand{\ljVary}{y}

\newcommand{\ljHandler}{h}

\newcommand{\ljFunction}{\lambda\ljVar.\ljExp}
\newcommand{\ljClosure}{f}
\newcommand\ljNoClosure{\Box} 

\newcommand{\ljExp}{e}
\newcommand{\ljExpf}{f}
\newcommand{\ljExpg}{g}

\newcommand{\ljLocation}{l}

\newcommand{\ljStore}{\sigma}

\newcommand{\ljEnv}{\rho}
\newcommand{\ljSEnv}{\widehat{\ljEnv}}
\newcommand\ljSEnvEmpty{\widehat\emptyset}

\newcommand{\ljTerm}{t}
\newcommand{\ljBer}{b}
\newcommand{\ljExn}{\bullet}

\newcommand{\con}{\ensuremath{\mathcal{C}}}

\newcommand{\conD}{\ensuremath{\mathcal{D}}}

\newcommand{\conI}{\ensuremath{\mathcal{I}}}
\newcommand{\conJ}{\ensuremath{\mathcal{J}}}

\newcommand{\conQ}{\ensuremath{\mathcal{Q}}}
\newcommand{\conR}{\ensuremath{\mathcal{R}}}

\newcommand{\conA}{\ensuremath{\mathcal{A}}}
\newcommand{\conM}{\ensuremath{\mathcal{M}}}

\newcommand{\conE}{\ensuremath{\mathcal{E}}}
\newcommand{\conF}{\ensuremath{\mathcal{F}}}

\newcommand{\defAbs}[2]{\Lambda#1.#2}
\newcommand{\defBase}[2]{\lambda#1.#2}
\newcommand{\defFun}[2]{#1\rightarrow#2}
\newcommand{\defDep}[2]{#1\rightarrow#2\,#1}
\newcommand{\defMap}[3]{#1[#2\mapsto#3]}
\newcommand{\defApp}[2]{#1(#2)}

\newcommand{\defAnd}[2]{#1\wedge#2}
\newcommand{\defOr}[2]{#1\vee#2}
\newcommand{\defNeg}[1]{\neg#1}

\newcommand{\defCap}[2]{#1\cap#2}
\newcommand{\defCup}[2]{#1\cup#2}

\newcommand{\defSet}[1]{\tilde#1}

\newcommand{\cbBlame}{\ell}
\newcommand{\cbVar}{\iota}
\newcommand{\cbIdent}{\flat}

\newcommand{\cb}{\kappa}

\newcommand\cbSolution{\mu}
\newcommand\cbSatisfies{\models}

\newcommand{\cbTrueL}{\syntax{true}}
\newcommand{\cbFalseL}{\syntax{false}}

\newcommand{\cbTrue}{\mathsf{t}}
\newcommand{\cbFalse}{\mathsf{f}}
\newcommand{\cbBot}{\bot}
\newcommand{\cbTop}{\top}

\newcommand{\cbLattice}{\mathds{B}}

\newcommand{\cbOrOp}{{\vee}}
\newcommand{\cbAndOp}{{\wedge}}
\newcommand{\cbNotOp}{{\neg}}
\newcommand{\cbImpOp}{{\Rightarrow}}

\newcommand{\cbState}{\varsigma}

\newcommand{\defCb}[2]{#1\blacktriangleleft#2}

\newcommand{\update}{\blacktriangleleft}

\newcommand\SubjectName{\textit{subject}} 
\newcommand\ContextName{\textit{context}} 

\newcommand{\subject}[1]{#1.\SubjectName} 
\newcommand{\context}[1]{#1.\ContextName} 

\newcommand{\cbMakeVal}[1]{\tau(#1)}

\newcommand{\C}{\mathcal{C}}

\newcommand{\OC}{\ensuremath{\mathcal{O}}}

\newcommand{\defFC}[2]{#1\rightarrow#2}
\newcommand{\defNot}[1]{\neg#1}

\newcommand{\RuleLjConst}{Const}

\newcommand{\RuleLjContractFresh}{Contract-Fresh}
\newcommand{\RuleLjContractSbx}{Contract-Sandbox}

\newcommand{\RuleLjConstructorFresh}{Constructor-Fresh}
\newcommand{\RuleLjConstructorSbx}{Constructor-Sandbox}

\newcommand{\RuleLjVar}{Var}

\newcommand{\RuleLjOp}{Op}
\newcommand{\RuleLjOpE}{Op-E}
\newcommand{\RuleLjOpF}{Op-F}

\newcommand{\RuleLjNew}{New}
\newcommand{\RuleLjNewE}{New-E}

\newcommand{\RuleLjGet}{Get}
\newcommand{\RuleLjGetPrototype}{Get-Prototype}
\newcommand{\RuleLjGetUndefined}{Get-Undefined}
\newcommand{\RuleLjGetE}{Get-E}
\newcommand{\RuleLjGetF}{Get-F}

\newcommand{\RuleLjPut}{Put}
\newcommand{\RuleLjPutE}{Put-E}
\newcommand{\RuleLjPutF}{Put-F}
\newcommand{\RuleLjPutG}{Put-G}

\newcommand{\RuleLjAbs}{Abs}

\newcommand{\RuleLjApp}{App}
\newcommand{\RuleLjAppE}{App-E}
\newcommand{\RuleLjAppF}{App-F}

\newcommand{\RuleLjAssert}{Assert}
\newcommand{\RuleLjAssertE}{Assert-E}
\newcommand{\RuleLjAssertF}{Assert-F}

\newcommand{\RuleLjConstruct}{Construct}
\newcommand{\RuleLjConstructF}{Construct-F}

\newcommand{\RuleLjBlame}{Blame}
\newcommand{\RuleLjNoBlame}{NoBlame}

\newcommand{\RuleLjWrapE}{Wrap-E}

\newcommand{\RuleAssertBase}{Assert-BaseContract}

\newcommand{\RuleAssertAnd}{Assert-AndContract}
\newcommand{\RuleAssertOr}{Assert-OrContract}
\newcommand{\RuleAssertNot}{Assert-NotContract}
\newcommand{\RuleAssertIntersection}{Assert-IntersectionContract}
\newcommand{\RuleAssertUnion}{Assert-UnionContract}
\newcommand{\RuleAssertDelayed}{Assert-DelayedContract}

\newcommand{\RuleAppFunction}{App-FunctionContract}
\newcommand{\RuleAppCap}{App-IntersectionContract}
\newcommand{\RuleAppOr}{App-OrContarct}
\newcommand{\RuleAppNeg}{App-NotContarct}
\newcommand{\RuleAppDependent}{App-DependentContract}
\newcommand{\RuleAppSandbox}{App-Sandbox}

\newcommand{\RuleGetContract}{Get-Contract}
\newcommand{\RuleGetNoContract}{Get-NoContract}
\newcommand{\RuleGetSandbox}{Get-Sandbox}

\newcommand{\RulePutNoContract}{Put-NoContract}
\newcommand{\RulePutContract}{Put-Contract}
\newcommand{\RulePutSandbox}{Put-Sandbox}

\newcommand{\RuleWrapConst}{Wrap-Constant}
\newcommand{\RuleWrapContract}{Wrap-Contract}
\newcommand{\RuleWrapConstructor}{Wrap-Constructor}

\newcommand{\RuleWrapNonProxy}{Wrap-NonProxyObject}
\newcommand{\RuleWrapExisting}{Wrap-Existing}
\newcommand{\RuleWrapProxy}{Wrap-Proxy}
\newcommand{\RuleWrapDelayed}{Wrap-Contract}

\newcommand{\RuleError}{-Error}

\newcommand{\RuleCSEmpty}{CS-Empty}
\newcommand{\RuleCSUnion}{CS-Union}

\newcommand{\RuleCTFlat}{CT-Flat}
\newcommand{\RuleCTFunction}{CT-Function}
\newcommand{\RuleCTIntersection}{CT-Intersection}
\newcommand{\RuleCTUnion}{CT-Union}
\newcommand{\RuleCTAnd}{CT-And}
\newcommand{\RuleCTOr}{CT-Or}
\newcommand{\RuleCTNeg}{CT-Negation}
\newcommand{\RuleCTSet}{CT-Set}

%% file: abstract.tex

\begin{abstract}
  \JSC{} is a language embedded, higher-order contract system for JavaScript which enforces
  contracts by run-time monitoring. Beyond providing the standard abstractions for building
  higher-order contracts (base, function, and object contracts), \JSC's novel contributions are its
  guarantee of non-interfering contract execution, its systematic approach to blame assignment, its
  support for contracts in the style of union and intersection types, and its notion of a
  parameterized contract scope, which is the building block for composable run-time generated
  contracts that generalize dependent function contracts.

  \JSC{} is implemented as a library so that all aspects of a contract can be specified using the
  full JavaScript language. The library relies on JavaScript proxies to guarantee full interposition
  for contracts. It further exploits JavaScript's reflective features to run contracts in a sandbox
  environment, which guarantees that the execution of contract code does not modify the application
  state. No source code transformation or change in the JavaScript run-time system is required.
  %
  The impact of contracts on execution speed is
  evaluated using the Google Octane benchmark.
\end{abstract}


%% file: body.tex

\section{Introduction}
\label{sec:introduction}

A contract specifies the interface of a software component by stating
obligations and benefits for the component's users.
Customarily contracts comprise invariants for objects and components
as well as pre- and postconditions for individual methods. 
Prima facie such contracts may be specified using straightforward assertions.
But further contract constructions are needed for contemporary
languages with first-class functions and other advanced abstractions.
These facilities require higher-order contracts as well as ways to
dynamically construct contracts that depend on run-time values.

Software contracts were introduced with Meyer's \emph{Design by
Contract}{\texttrademark} methodology \cite{Meyer1988} 
that stipulates the specification of contracts for all components
of a program and the monitoring of these contracts while the program
is running. Since then, the contract idea has taken off and systems 
for contract monitoring are available for many
languages~\cite{HinzeJeuringLoeh2006,jcontractor,Kramer1998,HelmHollandGangopadhyay1990,Chitil2012,FaehndrichBarnettLogozzo2010,CastagnaGesbertPadovani2009,Caminiti2012} and with a wealth of features~\cite{KeilThiemann2013-Proxy,HeideggerBieniusaThiemann2012-popl,BeloGreenbergIgarashiPierce2011,DisneyFlanaganMcCarthy2011,TovPucella2010,DimoulasPucellaFelleisen2009,AhmedFindlerSiekWadler2011}.
Contracts are particularly important for dynamically typed
languages as these languages only provide memory safety and dynamic
type safety.
Hence, it does not come as a surprise that the first 
higher-order contract systems were devised for Scheme and Racket
\cite{FindlerFelleisen2002}, out of the need to create
maintainable software. Other dynamic languages like
JavaScript\footnote{\url{http://kinsey.no/projects/jsContract/},
\newline\url{https://github.com/disnet/contracts.js}},
Python\footnote{\url{http://legacy.python.org/dev/peps/pep-0316/}},
Ruby\footnote{\url{https://github.com/egonSchiele/contracts.ruby}},
PHP\footnote{\url{https://github.com/wick-ed/php-by-contract}}, and
Lua\footnote{\url{http://luaforge.net/projects/luacontractor/}} have followed suit.

Many contract systems
\cite{HinzeJeuringLoeh2006,Chitil2012,FaehndrichBarnettLogozzo2010,Caminiti2012,KeilThiemann2013-Proxy,TovPucella2010,DimoulasPucellaFelleisen2009,AhmedFindlerSiekWadler2011}
are \emph{language-embedded}: contracts are first-class
values constructed through some library API.  This
approach is advantageous because it does not tie the contract system to a particular
implementation, it neither requires users to learn a separate contract language
nor do implementors have to develop specialized contract tools.
As the contract system can be distributed as a library, it is
easily extensible.

But there are also disadvantages because the
contract execution may get entangled with the application code. For
example, every contract system supports ``flat'' contracts which
assert that a predicate holds for a value. In most language-embedded systems, the
predicate is just a host-language function returning a boolean
value. Unlike a real predicate, such a function may have side effects
that change the behavior of the application code.


\paragraph*{Contributions}
\label{sec:contribution}
We present the design and implementation of \JSC{}, a language
embedded, higher-order contract system for 
JavaScript~\cite{ecmascript2009} which enforces contracts by run-time
monitoring. 
\JSC{} supports most features of existing systems and a range of novel
features that have not been implemented in this combination before.
No source code transformation or change in the JavaScript run-time
system is required. 
In particular, \JSC\ is the first contract system for JavaScript that
supports the standard features of contemporary contract systems
(embedded contract language,  JavaScript in flat contracts, contracts
as projections, full interposition using JavaScript proxies~\cite{CutsemMiller2010}) in
combination with the following three novel points.
\begin{enumerate}
  \item \emph{Noninterference.}
    Contracts are guaranteed not to exert side effects on a contract abiding
    program execution. A predicate is an arbitrary JavaScript
    function, which can access the state of the application program
    but which cannot change it. An exception thrown 
    by a predicate is not visible to the application program.
    Our guarantees are explained in detail in Section~\ref{sec:non-interference}.
  \item \emph{Dynamic contract construction.}
    Contracts can be constructed and composed at run time using
    contract abstractions \emph{without compromising
      noninterference}. A contract abstraction may contain 
    arbitrary JavaScript code; it may read from global state and it
    may maintain encapsulated local state. 
    The latter feature can be used to construct recursive
    contracts lazily or to remember values from the prestate of a function
    for checking the postcondition.
  \item \emph{New contract operators.}
    Beyond the standard contract constructors (flat, function, pairs),
    \JSC\ supports object, intersection, and union
    contracts. Furthermore, contracts can be combined arbitrarily with the boolean 
    connectives: conjunction, disjunction, and negation.
\end{enumerate}
The discussion of related work in Section~\ref{sec:related_work}
contains a detailed comparison with other systems.
The implementation of the system is available on the Web%
\footnote{\url{http://proglang.informatik.uni-freiburg.de/treatjs/}}.



\paragraph*{Overview}
\label{sec:overview}

The rest of this paper is organized as follows:
Section~\ref{sec:examples} introduces \JSC{} from a
programmer's point of view.
Section~\ref{sec:monitoring} specifies contract
monitoring and Section~\ref{sec:implementation} explains the 
principles underlying the implementation. Section~\ref{sec:evaluation}
reports our experiences from applying \JSC{} to a range of benchmark
programs. Section~\ref{sec:related_work} discusses related work and
Section~\ref{sec:conclusion} concludes. 


Further technical details, examples, and a formalization of contracts and contract
monitoring are included in the appendix.


\section{A \JSC{} Primer}
\label{sec:examples}

The design of \JSC\ obeys the following rationales.
\begin{itemize}
  \item \emph{Simplicity and orthogonality.}
    A core API provides the essential features in
    isolation. While complex contracts may require using the core API,
    the majority of contracts can be stated in terms of a convenience
    API that \JSC{} provides on top of the core.
  \item \emph{Non-interference.}
    Contract checking does not interfere with contract abiding executions of the host program.
  \item \emph{Composability.}
    Contracts can be composed arbitrarily.
\end{itemize}

A series of examples explains how contracts are
written and how contract monitoring works.
The contract API includes \emph{constructors} that build contracts from other
contracts and auxiliary data as well as an \lstinline{assert} function that
attaches a contract to a JavaScript value.

Our discussion focuses on general design issues for contracts and avoids
JavaScript specifics where possible. Contracts.js
\cite{Disney2013:contracts} provides contracts tailored to
the idiosyncrasies of JavaScript's object system---these may be added
to \JSC{} easily.



\subsection{Base Contracts}
\label{sec:base_contracts}

The base contract (aka flat contract) is the fundamental building block for all other
contracts. It is defined by a predicate and asserting it to a value
immediately sets it in action. We discuss it for completeness---all
contract libraries that we know of provide this functionality, but
they do not guarantee noninterference like \JSC{} does.

In JavaScript, any function can be used as a predicate, because any
return value can be converted to boolean. JavaScript programmers
speak of \emph{truthy} or \emph{falsy} about values that convert to
true or false. Thus, a predicate holds for a value if applying the 
function evaluates to a truthy value.

For example, the function \lstinline |typeOfNumber| checks
its argument to be a number. We apply the appropriate contract
constructor to create a base contract from it. 

\begin{lstlisting}[name=examples]
function |typeOfNumber| (arg) {
  return (typeof arg) === 'number';
};
var |typeNumber| = Contract.Base(|typeOfNumber|);*'\label{line:type-number}'*
\end{lstlisting}

\lstinline{|Contract|} is the object that encapsulates the \TJS{}
implementation. Its \lstinline{|assert|} function attaches a contract to 
a value. Attaching a base contract causes the predicate to be checked
immediately. If the predicate holds, \lstinline{|assert|} returns the 
original value. Otherwise, \lstinline{assert} signals a contract violation
blaming the \emph{subject}.

In the following example, the first \lstinline{assert} returns
\lstinline{1} whereas the second \lstinline{assert} fails.

\begin{lstlisting}[name=examples]
Contract.assert(1, |typeNumber|); // accepted
Contract.assert('a', |typeNumber|); // violation, blame subject 'a'
\end{lstlisting}

Listing~\ref{lst:utilities} defines a number of base contracts for
later use. Analogous to \lstinline{|typeNumber|}, the
contracts \lstinline|typeBoolean| and \lstinline{|typeString|} check the
type of their argument. 
Contract \lstinline{|isArray|} checks if the argument is an array.
Its correct implementation requires the \lstinline{With} operator,
which will be explained in Section~\ref{sec:contract-sandbox}.

\begin{lstlisting}[name=examples,float,caption={Some utility contracts.},label={lst:utilities}]
var |typeBoolean| = Contract.Base(function (arg) {
  return (typeof arg) === 'boolean';
});
var |typeString| = Contract.Base(function (arg) {
  return (typeof arg) === 'string';
});
var |isArray| = Contract.With({Array:Array},
  Contract.Base(function (arg) *'\label{line:is-array}'*{
    return (arg instanceof Array);
}));
\end{lstlisting}


\subsection{Higher-Order Contracts}
\label{sec:higher-order_contracts}

The example contracts in Subsection~\ref{sec:base_contracts}
are geared towards values of primitive type, but a base contract may also specify
properties of functions and other objects. 
However, base contracts are not sufficiently expressive to state
interesting properties of objects and functions.
For example, a contract should be able to express that a function maps
a number to a boolean or that a certain field access on an object always
returns a number. 


\subsubsection{Function Contracts}
\label{sec:function_contracts}

Following Findler and Felleisen \cite{FindlerFelleisen2002},
a function contract is built from zero or more contracts for the
arguments and one contract for the result of the function. Asserting the function
contract amounts to asserting the argument contracts to the arguments of each
call of the function and to asserting the result contract to the return value
of each call. Asserting a function contract to a value immediately
signals a contract violation if the value is not a
function. Nevertheless, we call a function contract \emph{delayed},
because asserting it to a function does not immediately signal a
contract violation. 

As a running example, we develop several contracts for the function
\lstinline{|cmpUnchecked|}, which compares two values and returns a boolean.

\begin{lstlisting}[name=examples]
function |cmpUnchecked|(x, y) {
  return (x > y);
}
\end{lstlisting}

Our first contract restricts the arguments to numbers and asserts that
the result of the comparison is a boolean.

\begin{lstlisting}[name=examples]
var |cmp| = Contract.assert(cmpUnchecked, 
              Contract.AFunction([|typeNumber|,|typeNumber|],|typeBoolean|));*'\label{line:sfun-num-num-bool}'*
\end{lstlisting}

\lstinline{AFunction} is the convenience constructor for a function contract. 
Its first argument is an array with $n$ contracts for the first $n$ function 
arguments and the last argument is the contract for the result. This contract 
constructor is sufficient for most functions that take a fixed number of arguments.

The contracted function accepts arguments that satisfies contract \lstinline{|typeNumber|}
and promises to return a value that satisfies \lstinline{|typeBoolean|}.
If there is call with an argument that violates its contract, then the function contract
raises an exception blaming the \emph{context}, which is the caller of the function that 
provides the wrong kind of argument. If the argument is ok but the result fails, then blame 
is assigned to the \emph{subject} (i.e., the function itself). Here are some examples that exercise 
\lstinline{cmp}.

\begin{lstlisting}[name=examples]
|cmp|(1,2); // accepted
|cmp|('a','b'); // violation, blame the context
\end{lstlisting}

To obtain a subject violation we use a broken version 
of \lstinline{cmpUnchecked} that sometimes returns a string.

\begin{lstlisting}[name=examples]
var |cmpBroken| = function(x, y) {
  return (x>0 && y>0) ? (x > y) : 'error';
}
var |faultyCmp| = Contract.assert(cmpBroken,
                    Contract.AFunction([|typeNumber|,|typeNumber|],|typeBoolean|));
|faultyCmp|(0,1); // violation, blame the subject
\end{lstlisting}

Higher-order contracts may be defined in the
usual way and their blame reporting in \JSC{} follows Findler and Felleisen
\cite{FindlerFelleisen2002}.
For example, a function \lstinline{sort}, which takes an array and a numeric comparison
function as arguments and which returns an array, may be specified by
the following contract, which demonstrates nesting of function contracts.

\begin{lstlisting}[name=examples]
var |sortNumbers| = Contract.AFunction([|isArray|, |cmp|], |isArray|);*'\label{line:sort}'*
\end{lstlisting}

Higher-order contracts open up new ways for a function not to fulfill its contract.
For example, \lstinline{sort} may violate the contract by calling its comparison function
(contracted with \lstinline{cmp})
with non-numeric arguments. Generally, the context is responsible to pass 
an argument that satisfies its specification to the function and to use the function's 
result according to its specification. Likewise, the function is responsible for the use 
of its arguments and in case the arguments meet their specification 
to return a value that conforms to its specification.

In general, a JavaScript function has no fixed arity and arguments are
passed to the function in a special array-like 
object, the \lstinline{arguments} object. Thus, the core
contract \lstinline{Function} takes two arguments. The first argument is an object contract 
(cf. Subsubsection~\ref{sec:object_contracts}) that
maps an argument index (starting from zero) to a contract. The second
argument is the contract for the function's return value.
Thus, \lstinline{AFunction} creates an object contract
from the array in its first argument and passes it to \lstinline{Function}.

Using the core \lstinline{Function} contract is a bit tricky because
it exposes the unwary contract writer to some JavaScript internals.
The contract \lstinline{Function(isArray, typeNumber)} checks whether
the arguments object is an array (which it is not), but it does \emph{not}
check the function's arguments. As a useful application of this feature, the following
contract \lstinline{twoArgs} checks that a function is called with exactly two arguments. 

\begin{lstlisting}[name=examples]
var |lengthTwo| = Contract.Base(function (args) *'\label{line:length-two}'*{
  return (args.length == 2);
});
var Any = Contract.Base (function() { return true; });
var twoArgs = Contract.Function(|lengthTwo|, |any|);*'\label{line:two-args}'*
\end{lstlisting}

\subsubsection{Object Contracts}
\label{sec:object_contracts}

Apart from base contracts that are checked immediately and delayed contracts for 
functions, \TJS{} provides contracts for objects.
An object contract is defined by a mapping from property names to contracts.
Asserting an object contract to a value immediately signals a
violation if the value is not an object. The contracts in the mapping
have no immediate effect.
However, when reading a property of the contracted object, the
contract associated with this property is asserted to the property
value. Similarly, when writing a property, the new value is 
checked against the contract. 
This way, each value read from a property and each value that is newly
written into the property is guaranteed to satisfy the property's contract.
Reads and writes to properties not listed in an object contract  are not checked.

The following object contract indicates that the \lstinline{length}
property of an object is a number.
The constructor \lstinline{AObject} expects the 
mapping from property names to contracts as a JavaScript object.

\begin{lstlisting}[name=examples]
var arraySpec = Contract.AObject({|length|:|typeNumber|});
\end{lstlisting}

Any array object would satisfy this contract. Each access to the \lstinline{length} property of the contracted array
would be checked to satisfy \lstinline{|typeNumber|}.

Blame assignment for property reads and writes is inspired by Reynolds
\cite{Reynolds1970} interface for a reference cell: each property is
represented as a pair of a getter and a setter function. Both, getter and
setter apply the same contract, but they generate different blame.
If the contract fails in the getter, then the \emph{subject} (i.e., the object) is blamed.
If the contract fails in the setter, then the \emph{context} (i.e.,
the assignment) is blamed.
The following example illustrates this behavior.

\begin{lstlisting}[name=examples]
var faultyObj = Contract.assert({length:'1'}, arraySpec);
faultyObj.length; // violation, blame the subject
faultyObj.length='1'; // violation, blame the context
\end{lstlisting}

An object contract may also serve as the domain portion in a function
contract. It gives rise to yet another equivalent way of writing the 
contract from Line~\ref{line:sfun-num-num-bool}.

\begin{lstlisting}[name=examples]
Contract.Function(
  Contract.AObject([|typeNumber|, |typeNumber|]), |typeBoolean|);
\end{lstlisting}

Functions may also take an intersection (cf. Section~\ref{sec:contract-combinators})
of a function contract and an object contract to address properties of functions
and \lstinline{this}. There is also a special \lstinline{Method} contract 
that includes a contract specification for \lstinline{this}.

%


\subsection{Combination of Contracts}
\label{sec:contract-combinators}

Beyond base, function, and object contracts, \JSC{} provides 
the intersection and union of contracts 
as well as the standard boolean operators on contracts: conjunction
(\lstinline{And}), disjunction (\lstinline{Or}), and 
negation (\lstinline{Not}). The result of an
operator on contracts is again a contract that may be further
composed.


For space reasons, we only discuss intersection and union contracts,
which are inspired by the corresponding operators in type theory.
If a value has two types, then we can assign it an
\emph{intersection type} \cite{CoppoDezani-Ciancaglini1978}.
It is well known that intersection types are useful to model
overloading and multiple inheritance. 

As an example, we revisit \lstinline{cmpUnchecked}, which we
contracted with \lstinline{cmpNumbers} in
Section~\ref{sec:function_contracts} to ensure that its arguments are
numbers. As the comparison operators are overloaded to work for
strings, too, the following contract is appropriate. 

\begin{lstlisting}[name=examples]
Contract.Intersection(
  Contract.AFunction([|typeNumber|, |typeNumber|], |typeBoolean|),
  Contract.AFunction([|typeString|, |typeString|], |typeBoolean|));
\end{lstlisting}

This contract blames the context if the contracted function is applied to 
arguments that fail both domain contracts, that is,
\lstinline{[typeNumber, typeNumber]} and
\lstinline{[typeString, typeString]}. 
The subject is blamed if a function call does not fulfill
the range contract that corresponds to a satisfied domain contract.

This interpretation coincides nicely with the meaning of an intersection type. 
The caller may apply the function to arguments both satisfying either
\lstinline{|typeNumber|} or \lstinline{|typeString|}. In general, the argument has to satisfy 
the union of \lstinline{|typeNumber|} and \lstinline{|typeString|}. For disjoint arguments the 
intersection contract behaves identically to the disjunction contract. 

As in type theory, the union contract is the dual of an intersection contract.
Exploiting the well-known type equivalence $(A\to C) \wedge (B \to C) = (A
\vee B) \to C$ \cite{BarbaneraDezani-CiancagliniLiguoro1995},
we may rephrase the above contract with a union contract, which  accepts either a pair of numbers or a pair
of strings as function arguments:
\begin{lstlisting}[name=examples]
Contract.Function(
  Contract.Union(
    Contract.AObject([|typeNumber|,  |typeNumber|]),
    Contract.AObject([|typeString|, |typeString|])),
 |typeBoolean|);
\end{lstlisting}

Next, we consider the union of two function contracts.
\begin{lstlisting}[name=examples]
var uf = Contract.Union(
  Contract.AFunction([typeNumber,typeNumber], |typeBoolean|),
  Contract.AFunction([typeString,typeString], |typeBoolean));
\end{lstlisting}

Asserting this contract severely restricts the domain of a function.
An argument is only acceptable if it is acceptable for all function contracts in the union. 
Thus, the context is blamed if it provides an argument that does not fulfill 
both constituent contracts. 
For example, \lstinline{uf} requires an
argument that is both a number and a string. As there is no such
argument, any caller will be blamed.

For a sensible application of a union of function contracts, the domains should overlap:

\begin{lstlisting}[name=examples]
Contract.Union(
  Contract.AFunction([|typeNumber|, |typeNumber|], |typeBoolean|),
  Contract.AFunction([|typeNumber|, |typeNumber|], |typeString|));
\end{lstlisting}

This contract is satisfied by a function that either always returns a
boolean value or by one that always returns an string value.
It is \emph{not} satisfied by a function that alternates between both
return types between calls. A misbehaving function is blamed on the first alternation.


\subsection{Sandboxing Contracts}
\label{sec:contract-sandbox}

All contracts of \JSC{} guarantee noninterference:
Program execution is not influenced by the evaluation of a terminating
predicate inside a base contract. That is, a program with contracts is
guaranteed to read the same values and write to the same objects as
without contracts. Furthermore, it either signals a contract violation
or returns a results that behaves the same as without contracts. 

To achieve this behavior, predicates must not write to
data structures visible outside of the predicate.
For this reason, predicate evaluation takes place in a sandbox that
hides all external bindings and places a write
protection on objects passed as parameters.

To illustrate, we recap the \lstinline{|typeNumber|} contract from Line~\ref{line:type-number}.
Without the sandbox we could  abstract the target type of
\lstinline{typeNumber} with a function and build base
contracts by applying the function to different
type names as in the following attempt:

\begin{lstlisting}[name=examples]
function badTypeOf(type) {
  return Contract.Base(function(arg) {*'\label{line:bad-type-of}'*
    return (typeof arg) === type;
  });
}
var |typeNumberBad|=badTypeOf('number');
var |typeStringBad|=badTypeOf('string');
\end{lstlisting}

However, this code fragment does not work as expected. The
implementation method for our sandbox reopens the closure of the
anonymous function in line~\ref{line:bad-type-of} and removes
the binding for \lstinline{type} from the contract's predicate.  Both
\lstinline{typeNumberBad} and \lstinline{typeStringBad} would be
stopped by the sandbox because they try to access the (apparently) global variable \lstinline{type}.
This step is required to guarantee noninterference, because the syntax of predicates is not restricted in 
their expressiveness and programmers may do arbitrary things,
including communicating via global variables or
modifying data outside the predicate's scope.

In general, read-only access to data (functions and objects)
is safe and many useful contracts (e.g., the \lstinline{isArray} contract from
Line~\ref{line:is-array} references the global variable
\lstinline{Array}) require access to global
variables, so a sandbox should permit regulated access. 

Without giving specific permission, the sandbox rejects \emph{any}
access to the \lstinline{Array} object and signals a sandbox violation.
To grant read permission, a new contract operator \lstinline{With}
is needed that makes an external reference available inside the
sandbox. The \lstinline{With} operator takes
a \emph{binding object} that maps identifiers to values and a
contract. Evaluating the resulting contract installs the binding
in the sandbox environment and then evaluates the constituent contract
with this binding in place. Each value passed into the sandbox (as an
argument or as a binding) is wrapped in an identity preserving 
membrane \cite{CutsemMiller2010} to ensure read-only access to the entire 
object structure.


The \lstinline{With} constructor is one approach
to build parameterized contracts by providing a form of dynamic binding.

\begin{lstlisting}[name=examples]
var |typeOf| = Contract.Base(function(arg) *'\label{line:typeof}'* {
  return (typeof arg) === type;
});
var |typeNumber|=Contract.With({type:'number'},|typeOf|);
var |typeString|=Contract.With({type:'string'},|typeOf|);
\end{lstlisting}



For aficionados of lexical scope, 
contract constructors, explained in the next subsection, are another means for
implementing parameterized contracts. 

%

\subsection{Contract Constructors}
\label{sec:contract-constructors}

While sandboxing guarantees noninterference, it prohibits the
formation of some useful contracts. For example, the
range portion of a function contract may depend on the arguments or
a contract may enforce a temporal property by remembering previous
function calls or previously accessed properties. Implementing such a facility requires that
predicates should be able to store data without affecting normal
program execution.

\JSC{} provides a \emph{contract constructor} \lstinline{Constructor}
for building a parameterized contract. The constructor takes a function that maps
the parameters to a contract.
This function is evaluated in a sandbox, like a predicate.
Unlike a predicate, the function may contain contract definitions and
must return a contract.
Each contract defined inside the sandbox is associated with the same
sandbox environment and shares the local variables and the parameters
visible in the function's scope.
No further sandboxing is needed for the predicates / base contracts
defined inside the sandbox.
The returned contract has no ties to the outside world and thus the included predicates 
will not be evaluated in the sandbox again. If such a predicate is called, the 
encapsulated sandbox environment can be used to store data for later use and without 
affecting normal program execution.

In the next example, a contract constructor builds a base contract 
from the name of a type. The constructor provides a lexically scoped alternative to the approach in Line~\ref{line:typeof}.

\begin{lstlisting}[name=examples]
var |Type| = Contract.Constructor(function(type) {
  return Contract.Base(function(arg) {
    return (typeof arg) === type;
  });
});
\end{lstlisting}

To obtain the actual contract we apply the constructor to parameters
with the method \lstinline{Contract.construct(|Type|, 'number')} or
by using the \lstinline{construct} method of the constructor.

\begin{lstlisting}[name=examples]
var |typeNumber| = |Type|.construct('number');
var |typeString| = |Type|.construct('string');
\end{lstlisting}


Let's consider yet another contract for a compare function. For this
contract, we only want the contract of the comparison to state that the two
arguments have the same type.

\begin{lstlisting}[name=examples]
Contract.Constructor(function() *'\label{line:ctor}'*{
  var type;
  var |getType| = Contract.Base(function (arg) {
    return type = (typeof arg);
  });
  var |checkType| = Contract.Base(function (arg) {
    return type === (typeof arg);
  });
  var typeBoolean = Contract.Base(function (arg) {
    return (typeof arg) === 'boolean';
  });
  return Contract.AFunction([getType, checkType], typeBoolean);
});
\end{lstlisting}

This code fragment defines a constructor with zero parameters
(viz. the empty parameter list in Line~\ref{line:ctor}). 
As there are no parameters, this example only uses the constructor to
install a shared scope for several contracts. The contract
\lstinline{|getType|} saves the type of the first argument. 
The comparison function has to satisfy a function contract
which compares the type of the second arguments with the saved type. 


\subsection{Dependent Contracts}
\label{sec:dependent-contracts}

A dependent contract is a contract on functions where the range
portion depends on the function argument.
The contract for the function's range can be created with a contract constructor.
This constructor is invoked with the caller's argument.
Additionally, it is possible to import pre-state values in the scope
of the constructor so that the returned contract may refer to those values.

\JSC{}'s dependent contract operation only builds a range contract in
this way; it does not check the domain as checking the domain may be
achieved by conjunction with another function contract. By either
pre- or postcomposing the other contract, the programmer may choose
between picky and lax semantics for dependent contracts (cf.\ \cite{GreenbergPierceWeirich2010}).

For example, a dependent contract \lstinline{PreserveLength}
may specify that an array processing function like \lstinline{sort} (Line~\ref{line:sort})
preserves the length of its input. The constructor receives the arguments (input array and
comparison function) of a function call and returns a contract
for the range that checks that the length of the input array is
equal to the length of the result. 

\begin{lstlisting}[name=examples]
var PreserveLength = Contract.Dependent(
  Contract.Constructor(function(input, cmp) {
    return Contract.Base(function (result) {
      return (input.length === result.length);
    });
}));
\end{lstlisting}

%
%
\section{Contract Monitoring}
\label{sec:monitoring}

This section explains how contract monitoring works
and how the outcome of a contract assertion is determined by 
the outcome of its constituents.
For space reasons we focus on the standard contract types 
(base, function, and object contracts) with intersection and union; we describe the boolean operators in
the supplemental material.


\subsection{Contracts and Normalization}
\label{sec:contracts}

A contract is either an immediate contract, a delayed contract, 
an intersection between an immediate and a delayed contract, or a union of contracts.
Immediate contracts may be checked right away when asserted to a value
whereas delayed contracts need to be checked later on.  
Only a base contract is immediate.

A delayed contract is a function contract, a dependent contract, an object contract, or
an intersection of delayed contracts. 
Intersections are included because all parts of an intersection must be checked on
each use of the contracted object: a call to a function or an access
to an object property.





The presence of operators like intersection and union has severe implications. In particular, 
a failing base contract must not signal a violation immediately because it may 
be enclosed in an intersection. Reporting the violation must be deferred until 
the enclosing operator is sure to fail.

To achieve the correct behavior for reporting violations,
monitoring \emph{normalizes} contracts before it starts contract enforcement.
Normalization separates the immediate parts of a contract from its
delayed parts so that each immediate contract can be evaluated
directly, whereas the remaining delayed contracts wrap the subject of
the contract in a proxy that asserts the contract when  
the subject is used.

To expose the immediate contracts,
normalization first pulls unions out of intersections by applying the
distributive law suitably.  The result is a union of intersections
where the operands of each intersection are either immediate contracts
or function contracts. At this point, monitoring can check all
immediate contracts and set up proxies for the remaining delayed
contracts. It remains to define the structure needed to implement
reporting of violations (i.e., blame) that is able to deal  with
arbitrary combinations of contracts. 


\subsection{Callbacks}
\label{sec:callbacks}

To assert a contract correctly, its evaluation must connect each contract
with the enclosing operations and it must keep track of the evaluation state of these
operations. In general, the signaling of a violation depends on a
combination of failures in different contracts.

This connection is modeled by so-called \emph{callbacks}. 
They are tied to a particular contract assertion and link each contract to its next enclosing 
operation or, at the top-level, its assertion. A callback linked to a
source-level assertion is called \emph{root callback}.
Each callback implements
a constraint that specifies the outcome of a contract assertion in terms of its constituents. 

A callback is implemented as a method that accepts the result of a contract assertion. The
method updates a shared property, it evaluates the constraint%
, and passes the result to the enclosing callback.


Each callback is related to one specific contract occurrence in the
program; there is at least  one
callback for each contract occurrence and there may be multiple
callbacks for a delayed contract
(e.g., a function contract). The callback is associated with a record that defines the blame assignment 
for the contract occurrence. This record contains two fields, $\SubjectName$ and $\ContextName$. 
The intuitive meaning of the fields is as follows. If the $\SubjectName$ field is false, then the 
contract fails blaming the subject (i.e., the value to which the contract is asserted). 
If the $\ContextName$ field is false, then the contract fails blaming the context 
(i.e., the user of the value to which the contract is asserted).

\subsection{Blame Calculation}
\label{sec:blame}

The fields in the record range over $\cbLattice_4$, the lattice underlying Belnap's
four-valued logic \cite{Belnap1977}, which is intended to deal with incomplete or 
inconsistent information. The set $\cbLattice_4=\{\cbBot, \cbFalse, \cbTrue, \cbTop\}$ of 
truth values forms a lattice modeling accumulated knowledge about the truth 
of a proposition. Thus, a truth value may be considered as the set of classical truth values
$\{\cbTrueL, \cbFalseL\}$ that have been observed so far for a proposition.
For instance, contracts are valued as $\cbBot$ before they are evaluated and $\cbTop$
signals potentially conflicting outcomes of repeated checking of the same contract.

As soon as a base contract's predicate returns, the contract's callback is applied
to its outcome. A function translates the outcome to a truth value according to JavaScript's 
idea of \emph{truthy} and \emph{falsy}, where $\ljFalse$, $\ljUndefined$, $\ljNull$, $\ljNaN$, and
$\ljEmyStr$ is interpreted as false. Exceptions thrown during
evaluation of a base contract are captured and count as $\cbTop$.

A new contract assertion signals a violation if a root callback maps any field to
$\cbFalse$ or $\cbTop$. Evaluation continues if only
internal fields have been set to $\cbFalse$ or $\cbTop$.

\subsection{Contract Assertion}
\label{sec:assertion}


Contract monitoring starts when calling the assert function with a value 
and a contract.

The top-level assertion first creates a new root callback that may signal
a contract violation later on and an empty sandbox object that serves
as the context for the internal contract monitoring. The sandbox object carries all 
external values visible to the contract.


Asserting a base contract to a value wraps the value
to avoid interference and applies the predicate to the wrapped value.
Finally, it applies the current callback function to the predicate's outcome.


Asserting a delayed contract to an object results in a proxy object
that contains the current sandbox object, the associated callback, and the contract itself.
It is an error to assert a delayed contract 
to a primitive value. 

Asserting a union contract first creates a new callback that combines
the outcomes of the subcontracts according to the blame propagation
rules for the union. Then it asserts the first subcontract (with a
reference to one input of the new callback) to the value before it asserts
the second subcontract (with a reference to the other input) to the resulting value.

Asserting a \lstinline{With} contract first wraps the values defined in the binding object 
and builds a new sandbox object by merging the resulting values and the current sandbox object.
Then it asserts its subcontract.

\subsection{Application, Read, and Assignment}
\label{sec:application}


Function application, property read, and property assignment distinguish two cases: either the 
operation applies directly to a non-proxy object or it applies to a proxy. If the target of the 
operation is not a proxy object, then the standard operation is applied.

If the target object is a proxy with a delayed contract, then the contract is checked when the 
object is used as follows.

A function application on a contracted function first creates a fresh callback that combines
the outcomes of the argument and range contract according to the propagation rules for functions.
Then it asserts the domain contract to the argument object with reference to the domain input of 
the new callback before it applies the function to the result. After completion, the range contract 
is applied to the function's result with reference to the range input of the callback.

A function application on a function contracted with a dependent contract first applies
the contract constructor to the argument and saves the resulting range
contract. Next, it applies the function to the argument and asserts
the computed range contract to the result.

A property access on a contracted object has two cases depending on the presence of a contract for
the accessed property. If a contract exists, then the contract is asserted to the value after
reading it from the target object and before writing it to the target object.
Otherwise, the operation applies directly to the target object. 

Property write creates a fresh callback that inverts the responsibility of the contract assertion 
(the context has to assign a value according to the contract).

An operation on a proxy with an intersection contract asserts the first subcontract to the value before 
it asserts the second subcontract to the resulting value. Both
assertions are connected to one input channel 
of a new callback that combines there outcomes according to the rules for intersection.

All contract assertions forward the sandbox object in the proxy to the subsequent
contract assertion.

\subsection{Sandbox Encapsulation}
\label{sec:sandboxing}

The sandbox ensures noninterference with the actual program code
by stopping evaluation of the predicate if it attempts to modify data that is visible outside the contract.%
\footnote{The sandbox cannot ensure termination of the predicate, of course.}

To ensure noninterference, the subject of a base contract, the argument of a dependent contract as
well as the value passed by a contract constructor are all wrapped in a membrane to ensure that the 
contract's code cannot modify them in any way.

To wrap a non-proxy object, the object is packaged in a fresh proxy along with the current sandbox object. 
This packaging ensures that further access to the wrapped object uses the current sandbox object.
If the object points to a contracted object, the wrap operation continues with the target object,
before adding all contracts from the contracted object. A primitive value is not wrapped. 

A property read on a sandboxed object forwards the operation to the target and wraps the
resulting behavior. An assignment to a sandboxed object is not allowed, thus it signals a sandbox violation.

The application of a sandboxed function recompiles the function by adding the given sandbox object 
to the head of the scope chain. Finally, it calls the function.
The function's argument and its result are known to be wrapped in this case.


\subsection{Contract Satisfaction}
\label{sec:satisfaction}

\tikzstyle{level 1}=[level distance=0.6cm, sibling distance=1cm]
\tikzstyle{level 2}=[level distance=1.2cm, sibling distance=3.2cm]
\tikzstyle{level 3}=[level distance=1.2cm, sibling distance=3.2cm]
\tikzstyle{level 4}=[level distance=1.2cm, sibling distance=1.6cm]

\begin{figure*}[t]
 \centering

\begin{minipage}[t]{0.475\textwidth}
 \centering

 \begin{tikzpicture}[<-,level/.style={sibling distance=180mm/#1}]
   \node  [] (top){}
   child {node [circle, draw] (root){$\bullet$}
   child {node [circle, draw] (1) {$\cap$}
   child {node [circle, draw] (1l) {$\rightarrow$}
   child {node [] (1ld) {\textit{Num}} }
   child {node [] (1lr) {\textit{Num}}}
 }
 child {node [circle, draw] (1r) {$\rightarrow$}
 child {node [] (1rd) {\textit{Str}}}
 child {node [] (1rr) {\textit{Str}}}
}
  }
};
\path  (1ld) -- (1l) node[draw=none, midway, left=4pt]{$(\cbTrue,\cbFalse)$};
\path  (1lr) -- (1l) node[draw=none, midway, right=4pt]{$(\cbTrue,\cbFalse)$};
\path  (1rd) -- (1r) node[draw=none, midway, left=4pt]{$(\cbTrue,\cbTrue)$};
\path  (1rr) -- (1r) node[draw=none, midway, right=4pt]{$(\cbTrue,\cbTrue)$};
\path  (1l) -- (1) node[draw=none, midway, left=4pt]{$(\cbFalse,\cbTrue)$};
\path  (1r) -- (1) node[draw=none, midway, right=4pt]{$(\cbTrue,\cbTrue)$};
\path  (1) -- (root) node[draw=none, midway, left=2pt]{$(\cbTrue,\cbTrue)$};
\path[draw,->]  (top)  (root) node[draw=none, midway, left=2pt]{$(\cbTrue,\cbTrue)$};
\end{tikzpicture}
%
%
\end{minipage}
\hfill
\vrule
\hfill
\begin{minipage}[t]{0.475\textwidth}
 \centering

\begin{tikzpicture}[<-,level/.style={sibling distance=180mm/#1}]
  \node  [] (top){}
  child {node [circle, draw] (root){$\bullet$}
  child {node [] (1) {$\vdots$}
}
child {node [circle, draw] (2) {$\cap$}
child {node [circle, draw] (2l) {$\rightarrow$}
child {node [] (2ld) {\textit{Num}}}
child {node [] (2lr) {\textit{Num}}}
  }
  child {node [circle, draw] (2r) {$\rightarrow$}
  child {node [] (2rd) {\textit{Str}}}
  child {node [] (2rr) {\textit{Str}}}
}
  }
};
\path  (2ld) -- (2l) node[draw=none, midway, left=4pt]{$(\cbTrue,\cbTrue)$};
\path  (2lr) -- (2l) node[draw=none, midway, right=4pt]{$(\cbTrue,\cbFalse)$};
\path  (2rd) -- (2r) node[draw=none, midway, left=4pt]{$(\cbTrue,\cbFalse)$};
\path  (2rr) -- (2r) node[draw=none, midway, right=4pt]{$(\cbTrue,\cbFalse)$};
\path  (2l) -> (2) node[draw=none, midway, left=4pt]{$(\cbTrue,\cbFalse)$};
\path  (2r) -> (2) node[draw=none, midway, right=4pt]{$(\cbFalse,\cbTrue)$};
\path  (1) -- (root) node[draw=none, midway, left=2pt]{$(\cbTrue,\cbTrue)$};
\path  (2) -- (root) node[draw=none, midway, right=2pt]{$(\cbTrue,\cbFalse)$};
\path[draw,->]  (top)  (root) node[draw=none, midway, left=2pt]{$(\cbTrue,\cbTop)$};
\end{tikzpicture}

\end{minipage}

\caption{Blame Calculation of $\mathit{addOne}=\lambda x.(x\!+\!'1')$ contracted with
$(\mathit{Num}\to\mathit{Num})\cap(\mathit{Str}\to\mathit{Str})$.
The left side shows the callback graph after applying $\mathit{addOne}$ to the 
string $\mathit{'1'}$ (first call). The right side shows the graph constructed after applying
$\mathit{addOne}$ to the number $\mathit{1}$ (second call).
Each node is a callback. Each edge an input channel.
The labeling next to the arrow shows the record $(\ContextName,\SubjectName)$.
The root callback ($\bullet$) collects the outcome of all delayed contract assertions.
}
\label{fig:blame}
\end{figure*}


The blame assignment for a function contract is calculated from the blame assignment for 
the argument and result contracts, which are available through the records
$\cbVar_d$ and $\cbVar_r$. A function does not fulfill a contract if it does not fulfill its obligations 
towards its argument $\context{\cbVar_d}$ \textbf{or} if the argument fulfills its contract, 
but the result does not fulfills its contract $\subject{\cbVar_d}\cbImpOp\subject{\cbVar_r}$.
The first part arises for higher-order functions, which may pass illegal arguments to their 
function-arguments. The second part corresponds to partial correctness of the function with 
respect to its contract. 

A function's context (caller) fulfills the contract if
it passes an argument that fulfills its contract $\subject{\cbVar_d}$ \textbf{and}
it uses the function result according to its contract $\context{\cbVar_r}$.
The second part becomes non-trivial with functions that return functions.

An object (subject) does not fulfill an object contract if a property access 
returns a value that does not fulfill the contract. An object's context (caller) does not 
fulfills the contract if it assigns an illegal value to a contracted property \textbf{or} it
does not uses the objects return according to its contract.

The outcome of read access on a contracted property $\subject{\cbVar_c}$ is directly related
to the parent callback and does not need a special constraint. A write
to a property guarded with contract $C$ generates blame like a
call to a function with contract $C \to \mathit{Any}$. (\textit{Any}
accepts any value.)

The blame assignment for an intersection contract is defined from its constituents at
$\cbVar_r$ and $\cbVar_l$. A subject fulfills an intersection contract
if it fulfills both constituent contracts: $\subject{\cbVar_r}
\cbAndOp \subject{\cbVar_l}$. A context, however, only needs to
fulfill one of the constituent contracts:  $\context{\cbVar_r}
\cbOrOp \context{\cbVar_l}$.

Dually to the intersection rule, the blame assignment for a union contract 
is determined from its constituents at $\cbVar_l$ and $\cbVar_r$. A subject fulfills a union contract
if it fulfills one of the constituent contracts: $\subject{\cbVar_l} \cbOrOp \subject{\cbVar_r}$. 
A context, however, needs to fulfill both constituent contracts: $\context{\cbVar_l}
\cbAndOp \context{\cbVar_r}$, because it does not known which contract is fulfilled by the subject. 




Figure~\ref{fig:blame} illustrates the working of callbacks.
After applying \lstinline{addOne} to \lstinline{'1'}, the first function contract 
($\mathit{Num}\to\mathit{Num}$) would fail blaming the context, whereas the second
contract ($\mathit{Str}\to\mathit{Str}$) succeeds. Because the context of an intersection may choose
which side to fulfill, the intersection is satisfied.

However, the second call which applies \lstinline{addOne} to \lstinline{1} raises an exception.
The first function contract fails, blaming the subject, whereas the second contract fails, 
blaming the context. Because the subject of an intersection has to fulfill both contracts, 
the intersection fails, blaming the subject.


\section{Implementation}
\label{sec:implementation}

The implementation is based on the JavaScript Proxy
API \cite{CutsemMiller2010,CutsemMiller2013}, a part of
the ECMAScript 6 draft standard. This API is implemented in Firefox 
since version 18.0 and in Chrome V8 since version 3.5. Our development
is based on the SpiderMonkey JavaScript engine.






\subsection{Delayed Contracts}
\label{sec:impl-delayed}


Delayed contracts are implemented using 
JavaScript Proxies \cite{CutsemMiller2010,CutsemMiller2013}, which guarantees full 
interposition by intercepting all operations.
The assertion of a delayed contract wraps the subject of the contract
in a proxy. The handler for the proxy contains the
contract and implements traps to mediate the use of the subject and to assert the contract.
No source code transformation or change in the JavaScript run-time system is required.





\subsection{Sandboxing}
\label{sec:impl-sandbox}

Our technique to implement sandboxing relies on all the evil and bad parts of JavaScript: the
\lstinline{eval} function and the \lstinline{with} statement. The basic idea is as follows. The
standard implementation of the \lstinline{toString} method of a user-defined JavaScript function returns a string
that contains the source code of that function. When \JSC{} puts a function (e.g., a predicate) in a sandbox, it first
\emph{decompiles} it by calling its \lstinline{toString} method. Applying \lstinline{eval} to the
resulting string creates a fresh variant of that function, \textbf{but} it dynamically rebinds the
free variables of the function to whatever is currently in the scope at the call site of
\lstinline{eval}.

JavaScript's \lstinline!with (obj) { ... body ... }! statement modifies the current environment by placing
\lstinline{obj} on top of the scope chain while executing
\lstinline{body}. With this construction, which is somewhat related to
dynamic binding \cite{HansonProebsting2001}, any property
defined in \lstinline{obj} shadows the corresponding binding deeper down in the scope chain. Thus,
we can add and shadow bindings, but we cannot remove them. Or can we?

It turns out that we can also abuse \lstinline{with} to \emph{remove} bindings! The trick is to wrap the new
bindings in a proxy object, use \lstinline{with} to put it on top of the scope chain, and to trap
the binding object's \lstinline{hasOwnProperty} method. When JavaScript traverses the scope chain to
resolve a variable reference \lstinline{x}, it calls \lstinline{hasOwnProperty(x)} on the objects of
the scope chain starting from the top. Inside the \lstinline{with}
statement, this traversal first checks the proxied binding object. If its \lstinline{hasOwnProperty} method always returns
true, then the traversal stops here and the JavaScript engine sends all read and write 
operations for free variables to the proxied binding object. This way,
we obtain full interposition and the 
handler of the proxied binding object has complete control over the
free variables in \lstinline{body}.

The \lstinline{With} contract is \JSC's interface to populate this binding object. The operators for
contract abstraction and dependent contracts all take care to stitch the code fragments together in
the correct scope. 
To avoid the frequent decompilation and \lstinline{eval} of the same
code, our implementation caches the compiled code
where applicable. 

No value is passed inside the sandbox without proper protection. Our
protection mechanism is inspired by 
\emph{Revocable Membranes} \cite{CutsemMiller2010,StricklandTobinHochstadtFindlerFlatt2012}.
A membrane serves as a regulated communication channel between two worlds, in this case between
an object/ a function and the rest of a program. A membrane is essentially a proxy that guards all
read operations and---in our case---stops all writes. If the result of a read operation is an
object, then it is recursively wrapped in a membrane before it is returned. 
Access to a property that is bound to a \lstinline{getter} function needs to decompile the 
\lstinline{getter} before its execution.
Care is taken to
preserve object identities when creating new wrappers (our membrane is \emph{identity preserving}).


We employ membranes to keep the sandbox apart from normal program execution thus guaranteeing
noninterference. In particular, we encapsulate objects
passed through the membrane, we enforce write protection, and we withhold external bindings 
from a function. 



\subsection{Noninterference}
\label{sec:non-interference}

The ideal contract system should not interfere with the execution of
application code. That is, as long as the application code does not
violate any contract, the application should run as if no contracts
were present. Borrowing terminology from security, this property is
called noninterference (NI) \cite{GoguenMeseguer1982}: with the
assumption that contract code runs 
at a higher level of security than application code, the low security application
code should not be able to observe the results of the high-level
contract computation.

Looking closer, we need to distinguish internal and external sources of
interference. Internal sources of interference arise from
executing unrestricted JavaScript code in the predicate of a
base contract. This code may try to write to an object that is visible to the
application, it may throw an exception, or it may not terminate.
Our implementation restricts all write operations to local objects using
sandboxing. It captures all exceptions and turns them into an
appropriate contract outcome. A timeout could be used to transform a
contract that may not terminate into an exception, alas, such a timeout
cannot be implemented in JavaScript.\footnote{%
  The JavaScript \lstinline{timeout} function only schedules a function to run when the currently
  running JavaScript code---presumably some event handler---stops. It cannot interrupt a running
  function.}

External interference arises from the interaction of
the contract system with the language. Two such issues arise in a
JavaScript contract system, exceptions and object equality.

Exceptions arise when a contract failure is encoded by a contract
exception, as it is done in Eiffel, Racket, and contracts.js. If an
application program catches exceptions, then it may become aware of
the presence of the contract system by observing an exception caused
by a contract violation. Our implementation avoids this problem by
reporting the violation and then using a JavaScript API method to quit
JavaScript execution\footnote{This aspect is customizable because the
  API method is not generally available. It can easily be overwritten to report 
a violation elsewhere or to throw an exception.}.

Object equality becomes an issue because function contracts as well as
object contracts are implemented by some kind of wrapper. The problem
arises if a wrapper is different (i.e., not pointer-equal)
from the wrapped object so that an equality test between wrapper and
wrapped object or between different wrappers for the same object (read: tests between object and
contracted object or between object with contract A and object with
contract B) in the application program returns false instead of true.

Our implementation uses JavaScript proxies to implement
wrappers. Unfortunately, JavaScript proxies are always different from
their wrapped objects and the only safe way to change that is by
modifying the proxy implementation in the JavaScript VM. See our
companion paper \cite{KeilThiemann2015-tproxy} for more discussion. There are proposals
based on preprocessing all uses of equality to proxy-dereferencing
equality, for example using SweetJS \cite{DisneyFaubionHermanFlanagan2014}, but they do not work in
combination with \texttt{eval} and hence do not provide full interposition.


\section{Evaluation}
\label{sec:evaluation}

This section reports on our experience with applying contracts to select programs.
We focus on the influence of contract assertion and sandboxing on the execution time.

All benchmarks were run on a machine with two AMD Opteron Processor 
with 2.20~GHz and 64~GB memory. All example runs and timings reported in this paper 
were obtained with the SpiderMonkey JavaScript engine.

\subsection{Benchmark Programs}
\label{sec:programs}

To evaluate our implementation, we applied it to JavaScript benchmark programs from 
the Google Octane 2.0 Benchmark Suite\footnote{\url{https://developers.google.com/octane/}}. 
%
Octane 2.0 consists of 17 programs that range from performance tests to
real-world web applications (Figure~\ref{fig:benchmark}), from an OS kernel simulation to
a portable PDF viewer. Each program focuses on a special purpose, for example, function and method
calls, arithmetic and bit operations, array manipulation, JavaScript
parsing and compilation, etc.

Octane reports its result in terms of a score. The Octane
FAQ\footnote{\url{https://developers.google.com/octane/faq}} 
explains the score as follows: ``\emph{In a nutshell: bigger is
better. Octane measures the time a test takes to complete and then assigns a score that is inversely
proportional to the run time.}'' The constants in this computation are chosen so that the current
overall score (i.e., the geometric mean of the individual scores) matches the overall score from
earlier releases of Octane and new benchmarks are integrated by
choosing the constants so that the geometric mean remains the
same. The rationale is to maintain comparability. 

\subsection{Methodology}
\label{sec:methodology}

To evaluate our implementation, we wrote a source-to-source compiler that first modifies the benchmark code
by wrapping each function expression\footnote{Function expressions are all expressions of the 
  form \lstinline{function(..) \{..\}}.} in an additional function. 
In a first run, this additional function wraps its target function in a proxy that, 
for each call to the function, records the data types of the arguments and of the function's return value.
This recording distinguishes the basic JavaScript data types \emph{boolean}, \emph{null}, \emph{undefined}, 
\emph{number}, \emph{string}, \emph{function}, and \emph{object}. 
Afterwards, the wrapper function is used to assert an appropriate function contract
to each function expression. These function contracts are built from 
the types recorded during the first phase. 
If more than one type is recorded at a given program point, then the
disjunction of the individual type contracts is generated.

All run-time measurements were taken from a deterministic run, which requires a predefined number
of iterations, and by using a warm-up run. 


\subsection{Results}
\label{sec:results}

\begin{figure*}[t]
  \centering
  \small
  \begin{tabular}{ l || r || r || r | r | r | r || r}
    \toprule
    \textbf{Benchmark}& 
    \textbf{F}& 
    \textbf{S}& 
    \textbf{w/o C}& 
    \textbf{w/o D}& 
    \textbf{w/o M}& 
    \textbf{w/o P}& 
    \textbf{B} 
    \\
    \midrule
    Richards& 
    0.391&
    0.519&
    0.582&
    0.782&
    0.781&
    0.903&
    11142\\
    DeltaBlue&
    0.276&
    0.360&
    0.409&
    0.544&
    0.544&
    0.625&
    17462\\
    Crypto&
    11888&
    12010&
    11912&
    11914&
    11986&
    11979&
    11879\\
    RayTrace&
    1.09&
    1.45&
    1.82&
    2.51&
    2.51&
    3.02&
    23896\\
    EarleyBoyer&
    5135&
    5292&
    5126&
    5205&
    5233&
    5242&
    5370\\
    RegExp&
    1208&
    1181&
    1205&
    1199&
    1212&
    1178&
    1207\\
    Splay&
    20.6&
    27.8&
    31.2&
    42.5&
    42.5&
    49.7&
    9555\\
    SplayLatency&
    73.1&
    99.7&
    109&
    151&
    151&
    177&
    6289\\
    NavierStokes&
    6234&
    7159&
    7924&
    9176&
    8943&
    9456&
    12612\\
    pdf.js&
    9191&
    9257&
    9548&
    9156&
    9222&
    9152&
    9236\\
    Mandreel&
    12555&
    12542&
    12586&
    12549&
    12346&
    12431&
    12580\\
    MandreelLatency&
    18741&
    18883&
    18741&
    18883&
    19027&
    18955&
    19398\\
    Gameboy Emulator&
    6.80&
    9.07&
    10.8& 
    14.9&
    14.9&
    17.7&
    23801\\
    Code loading &
    6245&
    6785&
    6937&
    7372&
    7335&
    7533&
    9324\\
    Box2DWeb&
    3.57&
    4.67&
    5.72&
    7.80&
    7.82&
    9.19&
    12528\\
    zlib&
    29108&
    28708&
    29025&
    29047&
    28926&
    29063&
    29185\\
    TypeScript&
    187&
    248&
    290&
    400&
    396&
    463&
    11958\\
    \bottomrule
  \end{tabular}
  \caption{%
    Scores for the Google Octane 2.0 Benchmark Suite (bigger is better).
    Column \textbf{F} (\emph{Full}) contains the scores for running with sandboxed contract assertion.
    Column \textbf{S} (\emph{System only}) contains the score values for running TreatJS without
    predicate evaluation (all predicates are set to \emph{true}) but with all internal components
    (callback, decompile, membrane).
    Column \textbf{w/o C} (\emph{without callback}) shows the scores
    from a full run (with predicates) but without callback updates.
    Column \textbf{w/o D} (\emph{without decompile}) shows the scores without recompiling functions.
    Column \textbf{w/o M} (\emph{without membrane}) lists the scores with contract assertion but
    without sandboxing (and thus without decompile).
    Column \textbf{w/o P} (\emph{without predicate}) shows the score values of raw contract
    assertions without predicate evaluation and thus without sandboxing, decompile, and callback
    updates.
    The last column \textbf{B} (\emph{Baseline}) gives the baseline scores without contract assertion.
  }
  \label{fig:benchmark}
\end{figure*}

\begin{figure*}[t]
  \centering
  \small
  \begin{tabular}{ l || r | r | r | r | r | r}
    \toprule
    \textbf{Benchmark} & 
    \multicolumn{3}{c |}{\textbf{Contract}} &
    \multicolumn{2}{c |}{\textbf{Sandbox}} &
    \textbf{Callback}
    \\ 
    \textbf{}&
    \textbf{A}&
    \textbf{I}&
    \textbf{P}&
    \textbf{M}&
    \textbf{D}&
    \textbf{}
    \\
    \midrule
    Richards& 
    24&
    1599377224&
    935751200&
    4678756000&
    4&
    3351504000\\
    DeltaBlue&
    54&
    2319477672&
    1340451212&
    6702256060&
    5&
    4744203248\\
    Crypto&
    1&
    5&
    3&
    15&
    3&
    13\\
    RayTrace&
    42&
    687240082&
    509234422&
    2546172110&
    4&
    2190186074\\
    EarleyBoyer&
    3944&
    89022&
    68172&
    340860&
    6&
    309120\\
    RegExp&
    0&
    0&
    0&
    0&
    0&
    0\\
    Splay&
    10&
    11620663&
    7067593&
    35337965&
    5&
    26231845\\
    SplayLatency&
    10&
    11620663&
    7067593&
    35337965&
    5&
    26231845\\
    NavierStokes&
    51&
    48334&
    39109&
    195545&
    5&
    177197\\
    pdf.js&
    3&
    15&
    9&
    45&
    4&
    39\\
    Mandreel&
    7&
    57&
    28&
    140&
    4&
    128\\
    MandreelLatency&
    7&
    57&
    28&
    140&
    4&
    128\\
    Gameboy Emulator&
    3206&
    141669753&
    97487985&
    487439925&
    5&
    399084085\\
    Code loading &
    5600&
    34800&
    18400&
    92000&
    4&
    70400\\
    Box2DWeb&
    20075&
    172755100&
    112664947&
    563324735&
    5&
    469141435\\
    zlib&
    0&
    0&
    0&
    0&
    0&
    0\\
    TypeScript&
    4&
    12673644&
    8449090&
    42245450&
    2&
    33796350\\
    \bottomrule
  \end{tabular}
  \caption{%
    Statistic from running the Google Octane 2.0 Benchmark Suite.
    Column \textbf{A} (\emph{Assert}) shows the numbers of top-level contract assertions.
    Column \textbf{I} (\emph{Internal}) contains the numbers of internal contract assertions
    whereby column \textbf{P} (\emph{Predicate}) lists the number of predicate evaluations. 
    Column \textbf{M} (\emph{Membrane}) shows the numbers of wrap operation and column \textbf{D}
    (\emph{Decompile}) show the numbers of decompile operations.
    The last column \textbf{Callback} gives the numbers of callback updates.
  }
  \label{fig:statistics}
\end{figure*}

Figure~\ref{fig:benchmark} contains the scores of all benchmark programs in different
configurations, which are explained in the figure's caption.
As expected, all scores decrease when adding contracts.
The impact of a contract depends on the frequency of its application.
A contract on a heavily used function (e.g., in \emph{Richards}, \emph{DeltaBlue}, or \emph{Splay}) 
causes a significantly higher decrease of the score.  
These examples show that the run-time impact of contract assertion depends on the program and on the
particular value that is monitored. While some programs like \emph{Richards},
\emph{DeltaBlue}, \emph{RayTrace}, and \emph{Splay} are heavily affected, others are almost unaffected:
\emph{Crypto}, \emph{NavierStokes}, and \emph{Mandreel}, for instance. 

In several cases the execution with contracts (or with a particular feature) is faster than without.
All such fluctuations in the score values are smaller than the standard deviation over several runs of the
particular benchmark.

For better understanding, Figure~\ref{fig:statistics} lists some numbers of internal counters. The
numbers indicate that the heavily affected benchmarks (\emph{Richards}, \emph{DeltaBlue},
\emph{RayTrace}, \emph{Splay}) contain a very large number of internal contract assertions.
Other benchmarks are either not affected (\emph{RegExp}, \emph{zlib}) or only slightly 
affected (\emph{Crypto}, \emph{pdf.js}, \emph{Mandreel}) by contracts.

For example, the \emph{Richards} benchmark performs 24 top-level contract assertions (these are
all calls to \lstinline{Contract.assert}), 
1.6 billion internal contract assertions (including
top-level assertions, \emph{delayed} contract checking, and predicate
evaluation), and 
936 million predicate executions. The sandbox wraps 
about 4.7 billion elements, but performs only 4 decompile
operations. Finally, contract checking performs 
3.4 billion callback update operations.

Because of the fluctuation in slightly affected benchmark programs the following discussion focuses
on benchmarks that were heavily impacted. Thus, we ignore the benchmark programs \emph{Crypto},
\emph{RegExp}, \emph{pdf.js}, \emph{Mandreel}, \emph{zlib}. 

In a first experiment, we turn off predicate execution and return \lstinline{true} instead of the predicate's
result. This splits the performance impact into the impact caused by the contract system (proxies,
callbacks, and sandboxing) and the impact caused by evaluating predicates. From the score values we find
that the execution of the programmer provided predicates causes a slowdown of 9.20\% over all
benchmarks (difference between \textbf{F} and \textbf{S}). The remaining slowdown is caused by the
contract system itself. The subsequent detailed treatment of the score values splits the impact into
its individual components.

Comparing columns \textbf{F} and \textbf{w/o C} shows that callback updates cause an overall
slowdown of 4.25\%. This point includes the recalculation of the
callback constraints as explained in Section~\ref{sec:satisfaction}.  

The numbers also show that decompiling functions has negligible impact on the execution time.
Decompiling decreases the score by 6.29\% over all benchmarks (compare columns \textbf{w/o C} and
\textbf{w/o D})\footnote{%
  Function recompilation can be safely deactivated for the benchmarks without 
  changing the outcome because our generated base contracts are
  guaranteed to be free of side effects.
}. 
This number is surprisingly low when taking into account that predicate evaluation includes
recompiling all predicates on all runs as explained in Section~\ref{sec:implementation}.

Comparing the scores in columns \textbf{w/o D} and \textbf{w/o M} indicates that the membrane, as
it is used by the sandbox, does not contribute significantly to the run-time overhead.
It does not decrease the total scores.

Finally, after deactivating predicate execution, we see that pure predicate handling causes a slowdown of
approximately 1.76\% (this is the impact of the function calls). In contrast to column \textbf{S},
column \textbf{w/o P} shows the score values of the programs without sandboxing, without
recompiling, and without callback updates, whereas in column \textbf{S} sandboxing, recompilation,
and callback updates remain active.

From the score values we find that the overall slowdown of sandboxed contract checking vs. a baseline
without contracts amounts to a factor of 7136, approximately. 
The dramatic decrease of the score values in the heavily affected benchmarks is simply caused by the
tremendous number of checks that arise during a run.

For example, in the \emph{Splay} benchmark, the  insert, find, and remove functions on trees are contracted. 
These functions are called every time a tree operation is performed. As the benchmark runs 
for 1400 iterations on trees with 8000 nodes, there is a considerable number of operations,
each of which checks a contract.
It should be recalled that every contract check performs at least two \lstinline{typeof} checks.


Expressed in absolute time spans, contract checking causes a run
time deterioration of 0.17ms for every single predicate check.
For example, the contracted \emph{Richards} requires 152480 seconds to complete and
performs 935751200 predicate checks. Its baseline requires 8 seconds. Thus, contract checking requires 
152472 seconds. That gives 0.16ms per predicate check.

Google claims that Octane ``measure[s] the performance of JavaScript code found in large, real-world
web applications, running on modern mobile and desktop
browsers''\footnote{\url{https://developers.google.com/octane/}}. For an academic, this is as
realistic as it gets. 

However, there are currently no large JavaScript application with contracts that we could use to benchmark,
so we had to resort to automatic generation and insertion of
contracts. These contracts may end up in artificial and unnatural
places that would be avoided by an efficiency conscious human developer. Thus, the numbers that we
obtain give insight into the performance of our contract implementation, but they cannot be used to
predict the performance impact of contracts on realistic programs with contracts inserted by human
programmers. The scores of the real-world programs (\emph{pdf.js}, \emph{Mandreel}, \emph{Code
  Loading}) among the benchmarks provide some initial evidence in this direction: their scores 
are much higher and they are only slightly effected by contract monitoring. But more experimentation
is needed to draw a statistically valid conclusion.


\section{Related Work}
\label{sec:related_work}

\paragraph*{Contract Validation}

Contracts may be validated statically or dynamically.
Purely static frameworks (e.g. ESC/Java \cite{FlanaganLeinoLillibridgeNelsonSaxeStata2002})
transform specifications and programs into verification conditions to be verified by a theorem
prover. Others \cite{XuJonesClaessen2009,Tobin-HochstadtHorn2012} rely on symbolic execution
to prove adherence to contracts. 
However, most frameworks perform run-time monitoring as proposed in Meyer's work.

\paragraph*{Higher-Order Contracts}

Findler and Felleisen \cite{FindlerFelleisen2002} first showed how to construct contracts and
contract monitors for higher-order functional languages. Their work has attracted a plethora of
follow-up works that range from semantic investigations \cite{BlumeMcallester2006,FindlerBlume2006}
over deliberations on blame assignment \cite{DimoulasFindlerFlanaganFelleisen2011,WadlerFindler2009}
to extensions in various directions:
contracts for polymorphic types \cite{AhmedFindlerSiekWadler2011,BeloGreenbergIgarashiPierce2011},
for affine types \cite{TovPucella2010}, 
for behavioral and temporal conditions
\cite{DimoulasTobin-HochstadtFelleisen2012,DisneyFlanaganMcCarthy2011},
etc. While the cited semantic investigations consider noninterference, only Disney and coworkers
\cite{DisneyFlanaganMcCarthy2011} give noninterference a high priority and propose an implementation
that enforces it. The other contract monitoring implementations that
we are aware of, do not address noninterference or restrict their predicates.

\paragraph*{Embedded Contract Language}

Specification Languages like JML \cite{LeavensBakerRuby1999} state behavior in terms of a custom
contract language or in terms of annotations in comments. An embedded contract language exploits the
language itself to state contracts. Thus programmers need not learn a new language and the contract
specification can use the full power of the language. Existing compilers and tools can be used
without modifications. 

\paragraph*{Combinations of Contracts}

Over time, a range of contract operators emanated, many of which are
inspired by type operators. There are
contract operators analogous to (dependent) function types \cite{FindlerFelleisen2002},
product types, sum types \cite{HinzeJeuringLoeh2006}, as well as
universal types \cite{AhmedFindlerSiekWadler2011}.
Racket also implements restricted versions of conjunctions and
disjunctions of contracts (see below).
However, current systems
do not support contracts analogous to union and intersection types
nor do they support full boolean combination of contracts (negation is
missing). 

Dimoulas and Felleisen \cite{DimoulasFelleisen2011} propose a contract composition, 
which corresponds to a conjunction of contracts. But their operator is
restricted to contracts  of the same type. Before evaluating a conjunction it lifts the operator recursively to 
base contracts where it finally builds the conjunction of the predicate results.

Racket's contract system \cite[Chapter 7]{FlattFindlerPlt2014:_racket_guide} supports
boolean combinations of contracts. Conjunctions of contracts are decomposed and 
applied sequentially \cite{Tobin-HochstadtHorn2012}. Disjunctions of flat contracts 
are transformed so that the first disjunct does not cause a blame immediately if 
its predicate fails. However, Racket places severe restrictions on using disjunction with 
higher-order contracts and restricts negation to base contracts. A
disjunction must be resolved by first-order choice to at most one
higher-order contract; otherwise it is rejected at run time.


\paragraph*{Proxies}

The JavaScript proxy API \cite{CutsemMiller2010} enables a developer to enhance the functionality of
objects easily. JavaScript proxies have been used for Disney's
JavaScript contract system, contracts.js \cite{Disney2013:contracts}, to enforce access permission contracts
\cite{KeilThiemann2013-Proxy}, as well as for other dynamic effects systems, meta-level extension,
behavioral reflection, security, or concurrency control
\cite{MillerCutsemTulloh2013,AustinDisneyFlanagan2011,BrachaUngar2004}.



\paragraph*{Sandboxing JavaScript}

The most closely related work to our sandbox mechanism is the work of Arnaud et al.
\cite{ArnaudDenkerDucassePolletBergelSuen2010}. They provide features similar to our sandbox
mechanism. Both approaches focus on access restriction and noninterference to guarantee side effect
free assertions of contracts.

Our sandbox mechanism is inspired by the design of access control wrappers which is used for
revocable references and membranes \cite{CutsemMiller2010,Miller2006}. In memory-safe languages, a
function can only cause effects to objects outside itself if it holds a reference to the other
object. The authority to affect the object can simply be removed if a membrane revokes the reference
which detaches the proxy from its target.

Our sandbox works in a similar way and guarantees read-only access to target objects, but redirects
write operations. Write access is completely forbidden and raises an exception. However, the
restrictions affect only values that cross the border between the global execution environment and a
predicate execution. Values that are defined and used in one side, e.g. local values, were not
restricted. Write access to those values is fine.

Other approaches implement restrictions by filtering and rewriting untrusted code or by removing 
features that are either unsafe or that grant uncontrolled access.
The Caja Compiler \cite{GoogleCaja,MillerSamuelLaurieAwadStay:caja_safe}, for example, 
compiles JavaScript code in a sanitized JavaScript subset that can safely be executed in normal 
engines.
However, some static guarantees do not apply to code created at run time. 
For this reason Caja restricts dynamic features and adds run-time checks that 
prevent access to unsafe function and objects.



\section{Conclusion}
\label{sec:conclusion}

We presented \JSC{}, a language embedded, dynamic, higher-order contract system for full JavaScript.
\JSC{} extends the standard abstractions for higher-order contracts with intersection and union contracts,
boolean combinations of contracts, and parameterized contracts, which are the building blocks for contracts 
that depend on run-time values. 
\JSC{} implements proxy-based sandboxing for all code fragments in contracts to
guarantee that contract evaluation does not interfere with normal program execution. 
The only serious impediment to full noninterference lies in JavaScript's treatment of proxy
equality, which considers a proxy as an individual object. 

The impact 
of contracts on the execution time
varies widely depending on the particular functions that are under contract and on the frequency 
with which the functions are called. 
While some programs' run time is heavily impacted, others are
nearly unaffected. We believe that if contracts are carefully and manually inserted with the purpose
of determining interface specifications and finding bugs in a program, their run time will mostly be
unaffected. But more experimentation is needed to draw a statistically valid conclusion.


%% file: appendix.tex
\input{sm.body.examples}
\input{sm.body.syntax}
\input{sm.body.monitoring}

%% file: sm.body.examples.tex
\section{Contracts in a Higher-order World}

In addition to contracts already mentioned in the paper, \TJS{} offers a wealth of further features included in 
its core or convenience API.

Lets start of with defining a number of base contracts for later use.
\begin{lstlisting}[name=examples2]
var |typeNumber| = Contract.Base(function (arg) {
  return (typeof arg) === 'number';
});
var |typeBoolean| = Contract.Base(function (arg) {
  return (typeof arg) === 'boolean';
});
var |typeString| = Contract.Base(function (arg) {
  return (typeof arg) === 'string';
});
var |isArray| = Contract.With({Array:Array}, Contract.Base(function (arg) *'\label{line:is-array}'*{
  return (arg instanceof Array);
}));
var |isUndefined| = Contract.Base(function (arg) {
  return (arg === undefined);
});
var |Any| = Contract.Base(function (arg) {
  return true;
});
\end{lstlisting}
Analogous to \lstinline{|typeNumber|}, the
contracts \lstinline{|typeBoolean|} and \lstinline{|typeString|} check the
type of their argument. 
Contract \lstinline{|isArray|} checks if the
argument is an array, \lstinline{|isUndefined|}
checks for \lstinline{undefined}, and \lstinline{|Any|} is a contract
that accepts any value.

\subsection{Simple Function Contracts}

To keep the contract system usable it is important that contract definitions are as short as possible.
So, instead of defining a function contract in terms of an object contract, the simplest way
to define a function contract is to give a sequence of contracts.

The function \lstinline{SFunction} is the constructor for a \emph{simple} function contract. 
Its arguments are contracts. If called with $n+1$ arguments for $n\ge 0$, then the first $n$ 
arguments are contracts for the first $n$ function arguments and the last
argument is the contract for the result. This contract constructor is
sufficient for most functions that take a fixed number of arguments.

The following function contract is another simple function contract equivalent to the
contract already defined in the paper (Line~\ref{line:sfun-num-num-bool}).
The constructor function \lstinline{SFunction} implicitly creates an object contract
from its $n$ arguments.

\begin{lstlisting}[name=examples2]
Contract.assert(cmpUnchecked, Contract.SFunction(|typeNumber|, |typeNumber|, |typeBoolean|));
\end{lstlisting}

\subsection{Object and Method Contracts}

A combination of object and function contracts is useful for
specifying the methods of an object. The contract below is a fragment
of such a specification. It states that the \lstinline{length}
property is a number and that the \lstinline{foreach} property is a
function that takes a number and any value to return \lstinline{undefined}.
The function stored in \lstinline{foreach} is to
return \lstinline{undefined}.\footnote{Recall that \lstinline{undefined} is a proper value
in JavaScript. All functions that have no return value in fact return \lstinline{undefined}.}

\begin{lstlisting}[name=examples2]
Contract.AObject({
  |length|:|typeNumber|,
  |foreach|:Contract.AFunction([Contract.AFunction([|typeNumber|, |Any|], |isUndefined|)], |isUndefined|)
});
\end{lstlisting}

Apart from normal function contracts, \JSC{} provides a \emph{method contract}. A method contract extends 
a function contract by a third portion, applied to the \lstinline{this} reference. A similar extension 
is applies to dependent contracts.

\begin{lstlisting}[name=examples2]
Contract.AMethod([Contract.AFunction([|typeNumber|,|Any|],|isUndefined|)],|isUndefined|,isArray);
\end{lstlisting}

This fragment defined a contract for a \lstinline{foreach} method of an \lstinline{array} object. 
The first two arguments are equivalent to the definition of a function contract. The third portion 
indicates that the \lstinline{this} reference has to be an instance of \lstinline{Array}.

The following contract now combines both contracts.

\begin{lstlisting}[name=examples2]
Contract.AMethod([Contract.AFunction([|typeNumber|,|Any|],|isUndefined|)],|isUndefined|, Contract.AObject({|length|:|typeNumber|}));
\end{lstlisting}

\subsection{Regular Expression Matching}

An object contract maintains a mapping from property names to contracts. But, this mapping is inconvenient 
if more than one property match to the same contact. To simplify this, \JSC{} also provides regular expression 
matching in object contracts.

Core object contracts are drawn from a mapping, that points property names to contracts. This mapping can either be a
string map, which maps strings to contracts, or a regular expression map that contains pairs of a regular expression 
and a contract. When accessing a property the map checks the presents of the property's name and 
returns the set of associated contracts.

\begin{lstlisting}[name=examples2]
Contract.Object(Contract.Map.RegExMap([Contract.Map.Mapping(/.*/, typeNumber)]));
\end{lstlisting}

\subsection{Reflection Contracts}

Object contracts follow Reynolds \cite{Reynolds1970} interface for a reference cell: each property is
represented as a pair of a getter and a setter function. Both, getter and setter
apply the same contract but on different portions.

However, property access and property assignment can be seen as a meta-level
call. A \lstinline{get} function accepts a target value and a property name and
returns a value, whereas a \lstinline{set} function accepts target object, a
property name, and a value and
returns another value. This corresponds to JavaScrip's get and set trap as
specified in its reflection API. 

\TJS\ enables to define reflection contract which apply directly to the
corresponding operation. In detail, a reflection contract requires a function
contract as argument which is applied to the corresponding trap.

The following examples demonstrated an get contract that checks whether the
accessed property exists or not.
\begin{lstlisting}[name=examples2]
var PropertyCheck = Contract.Constructor(function() {
  var target = {};
  var getTraget = Contract.Base(function(arg) {
    return target=arg;
  });
  var checkProperty = Contract.Base(function(name) {
    return (name in target);
  });
  var any = Contract.Base(function() {
    return true;
  });
  return Contract.Get(Contract.AFunction([getTraget, checkProperty], any));
});
\end{lstlisting}
The constructor first defines a base contract that sores the target object,
whereas the second base contracts checks for an undefined property. 

Other reflection contracts can be build in the same way

\subsection{Boolean Combination of Contracts}

Beyond base, function, and object contracts, \JSC{} provides the
standard boolean operators on contracts: conjunction
(\lstinline{And}), disjunction (\lstinline{Or}), and 
negation (\lstinline{Not}) with the obvious meanings. The result of a boolean
operator on contracts is again a contract that may be further
composed.

The following example demonstrates boolean operators with base
contracts. The contracts should be self-explanatory and none of them
should fail.

\begin{lstlisting}[name=examples2]
var |isPositive| = Contract.Base(function (arg) {
  return (arg > 0);
});
Contract.assert(1, Contract.And(|typeNumber|, |isPositive|));
Contract.assert('k', Contract.Or(|typeNumber|, |typeString|));
Contract.assert(1, Contract.Not(|isUndefined|));
\end{lstlisting}

Boolean operators are not really exciting for base contracts as they
could be pushed into the predicates. However, their use
improves modularity and they open interesting possibilities in
combination with function and object contracts.

For example, we may combine the \lstinline{lengthTwo} contract
with the contract on \lstinline{cmp} to further sharpen the contract for
a function like \lstinline{cmp}. This contract guarantees that a
function is only ever invoked with exactly two arguments, that these
arguments must be numbers, and that the function must return a boolean.

\begin{lstlisting}[name=examples2]
var |lengthTwo| = Contract.Base(function (args) {
  return (args.length == 2);
});
var |cmpNumbers| = Contract.AFunction([|typeNumber|, |typeNumber|], |typeBoolean|)
Contract.And (lengthTwo, |cmpNumbers|);
\end{lstlisting}

Boolean operators may also express alternatives.
A contract for \lstinline{cmp} that reflects that JavaScript's $<$
operator also compares strings would look as follows.

\begin{lstlisting}[name=examples2]
var |cmpStrings| = Contract.AFunction([|typeString|, |typeString|], |typeBoolean|);
Contract.Or(|cmpNumbers|, |cmpStrings|);
\end{lstlisting}

The disjunction of two function contracts requires that each call site fulfills one
of the constituent contracts. As both functions contracts promises to return a value 
that satisfies \lstinline{|typeBoolean|} we can alternately lift the disjunction 
to the arguments.

\begin{lstlisting}[name=examples2]
Contract.Function(Contract.Or(Contract.AObject([|typeNumber|, |typeNumber|]), Contract.AObject([|typeString|, |typeString|])), |typeBoolean|);
\end{lstlisting}

Using the conjunction of these two function
contracts would always signal a violation because no argument fulfills
\lstinline{|typeNumber|} and \lstinline{|typeString|} at the same time.

For conjunctions blaming is easy because all parts of the conjuncted contracts
have to hold. For disjunctions the context is to blame if it does not satisfy one 
of the disjuncts. If at least one domain portion fits but the results does hold for 
one of the corresponding range contracts blame is assigned to the subject.

At first glance, a negated function contract seems to be pointless,
because negation could be eliminated by pushing it into the
predicates of the base contracts. But, again, there are modularity and
reusability benefits and the use of negation can give rise to succinct
contracts.

Consider the contract \lstinline{|typeNumber|}$\rightarrow$\lstinline{|isPositive|}.
Asserting it to a function requires that \lstinline{|typeNumber|}
accepts the argument value and \lstinline{|isPositive|} accepts
the return value. Otherwise, a contract violation is signaled.
However, the programmer may have a much weaker guarantee in mind. The
only useful guarantee may be that if the argument satisfies
\lstinline{|typeNumber|}, then \lstinline{|isPositive|} accepts the
return value. A negated function contract expresses such a (weak)
guarantee succinctly.

\begin{lstlisting}[name=examples2]
Contract.Not(Contract.SFunction(|typeNumber|, Contract.Not(|isPositive|)));
\end{lstlisting}

The inner contract accepts only if the argument is a number, but the
return value is not greater than zero. Thus, its negation has the
desired behavior. Of course, such a contract could be written without
a negated function contract, but it is much less perspicuous.

\begin{lstlisting}[name=examples2]
Contract.Or(Contract.SFunction(|typeNumber|, |isPositive|), Contract.SFunction(Contract.Not(|typeNumber|), |Any|));
\end{lstlisting}

However, the possibility to compose contracts arbitrarily without restriction 
enables a programmer to build higher-level connections in top of the boolean combinators
e.g. a conditional (implication), a biconditional, a exclusive
disjunction, a alternative denial, and a joint denial.

\subsection{Weak and Strict Contracts}

Let's recapitulate the two kinds of contracts assigned to the \lstinline{cmp} function.

\begin{lstlisting}[name=examples2]
Contract.And (lengthTwo, |cmpNumbers|);
\end{lstlisting}

The first one checks that the function is called with exactly two arguments, where the second indicates 
that both arguments are of type number. This combination is required to get a \emph{strict} 
interpretation of the specified contract. 

Now, we have to admit that a function contract like \lstinline{Contract.SFunction(IsNumber, IsNumber, IsBoolean)} 
gives a \emph{weak} interpretation. If the contracted function only takes use of the first argument, the second 
would never be checked. This makes it possible that the function is called with a pair of type 
$(\mathit{Num}, \mathit{String})$ without raising a contract violation. This is because the domain contracts 
are mapped to the arguments array whose check is only when accessing an element (similar to an object contract).

In JavaScript it is possible to call a function with more or less arguments then specified in a function. Each argument 
can be addressed by accessing the arguments array directly. Because the given interpretation is neither right nor wrong, 
\JSC{} makes it possible to differ between a \emph{weak} and a \emph{strict} interpretation of object contracts. 
This can be done by setting a flag when defining the object contract.

\begin{lstlisting}[name=examples2]
var weak = Contract.AObject({a:typeNumber,b:typeString,c:typeBoolean}, false);
var strict = Contract.AObject({a:typeNumber,b:typeString,c:typeBoolean}, true);
\end{lstlisting}

Forcing the object contract to be strict, each specified property gets immediately checked 
when the contract is asserted.

\subsection{Sandboxing Contracts}

Consider the following base contract where the programmer carelessly
omitted the \lstinline{var} keyword in the predicate.

\begin{lstlisting}[name=examples2]
var |faultyLengthTwo| = Contract.Base(function (arg) {
  length = arg.length;
  return (length===2);
});
\end{lstlisting}

If we assert \lstinline{|faultyLengthTwo|} to a value, the
predicate attempts to write to the global variable
\lstinline{length}. But the sandbox intercepts this write operation
and throws an exception. 

Passing a function into the sandbox poses a special problem because
the function may perform side effects on its free variables. 
The sandbox removes these free variable bindings so that
any use of \lstinline{faultyLengthTwo} signals a violation.

\subsection{Parameterized Contracts}

Because sandboxing restricts any access to the outside world, 
\TJS{} provides a \lstinline{With} contracts that binds values and a contract 
constructor that defines contracts apart from the normal program execution.

This setup enables us to build parameterized contracts by substituting parameters inside 
of base contracts.

Lets recap the \lstinline{|lengthTwo|} contract.
Using \lstinline{With} enables to exclude the targets length and to reuse the 
base contract with various objective.

\begin{lstlisting}[name=examples2]
var |hasLength| = Contract.Base(function(arg) *'\label{line:two-args2e}'* {
  return (arg.length===length);
});
var |lengthTwo| = Contract.With({length:2}, |hasLength|);
var |lengthThree| = Contract.With({length:3}, |hasLength|);
\end{lstlisting}

The following example shows a contract constructor that builds a base contracts to 
which checks the length property of its argument. The constructor gives an alternative
notation to the contract in Line~\ref{line:two-args2e}.

\begin{lstlisting}[name=examples2]
var |Length| = Contract.Constructor(function(length) *'\label{line:length-e}'*{
  return Contract.Base(function(arg) {
    return (arg.length===length);
  });
});
\end{lstlisting}

To obtain the actual contract we need to apply the constructor to some parameters
with the method \lstinline{Contract.construct(|Length|, 2)} or
we gather the constructor's construct method, which is referred by its \lstinline{construct}
property, and use this as a parameterisable contract.

\begin{lstlisting}[name=examples2]
var |LengthTwo| = |Length|.construct(2);
var |LengthThree| = |Length|.construct(3);
\end{lstlisting}

The constructor function takes one parameter, \lstinline{length}, builds a base contract 
that checks the \lstinline{length} property of an object, and returns an appropriate contract.

\subsection{Contract Abstraction}

Let's consider another use of constructors, contract abstraction. Contract
abstraction happens if one asserts a constructor directly to a value instead of
applying a value and deriving a contract from it.

Let's recap the \lstinline{cmp} function described in the paper. The function
may accept values of type number, string, or boolean as argument and guarantees
to return a boolean value. Instead of building the intersection between
different alternatives one can build an contract abstraction that first needs to
be applied with a value that specifies the input's type.

First, lets define an appropriate constructor.
\begin{lstlisting}[name=examples2]
var |Cmp| = Contract.Constructor(function(type) *'\label{line:cmp-ctor}'*{
  var typeOf = Contract.Base(function(arg) {
    return type === (typeof arg);
  });
  var typeBoolean = Contract.Base(function(arg) {
    return (typeof arg) === 'boolean';
  });
  return Contract.AFunction([typeOf, typeOf], typeBoolean);
});
\end{lstlisting}
The contractor abstracts the input type and returns a function contract with
respect to the constructor's argument. 

Instead of building a contract from such a constructor, one can assert the
contract to a value. Constructors are contracts and can be asserted to values as
contracts are.
\begin{lstlisting}[name=examples2]
var |cmpAbs| = Contract.assert(cmp, Cmp);
\end{lstlisting}
\lstinline{cmpAbs} is an abstraction of the \lstinline{cmp} function. To
unroll the abstraction we have to call \lstinline{cmpAbs} with the arguments
required by the constructor.
\begin{lstlisting}[name=examples2]
var |cmpNumber| = |cmpAbs|('number');
var |cmpString| = |cmpAbs|('string');
\end{lstlisting}
Here, \lstinline{cmpNumber} (\lstinline{cmpString}) is a contracted versions of
\lstinline{cmp} that accepts number (string) values as argument.
The following code snipped shows how to use the abstraction in one step.
\begin{lstlisting}[name=examples2]
var |result| = |cmpAbs|('number')(1,2);
\end{lstlisting}
Instead of passing around type information, constructors can abstract over
contracts. The following example shows another abstraction for \lstinline{cmp}.
\begin{lstlisting}[name=examples2]
var |Cmp2| = Contract.Constructor(function(typeOf) *'\label{line:cmp-ctor2}'*{
  var typeBoolean = Contract.Base(function(arg) {
    return (typeof arg) === 'boolean';
  });
  return Contract.AFunction([typeOf, typeOf], typeBoolean);
});
var |cmpAbs2| = Contract.assert(cmp, Cmp2);
var |result| = |cmpAbs2|(typeNumber)(1,2);
\end{lstlisting}

\subsection{Recursive Contracts}

Yet another use of contract constructors is recursion. Recursive contracts are
similar to constructors, but instead of building an abstraction the constructor
function is pending and evaluates to a contract when asserted. Then, the
recursive contract is given as argument to the constructor function.

The following example defines a \emph{recursive contract} for a linked list 
so that the \lstinline{next} property satisfies the same contract as the current element.
\begin{lstlisting}[name=examples2]
var LinkedListCtor = Contract.SRecursive(function ctorFun(recursive) {
  return Contract.Object({
    val:IsNumber,
    next:recursive
  });
});
\end{lstlisting}
The code fragment defines a constructor based on a named function \lstinline{ctorFun}. Each read access to 
the \lstinline{next} property of a contracted object would assert a new instance of itself to return a contracted object.

\subsection{Consistency of Values}

A contract constructor may also verify that a value does not change during a function call. The following example illustrates how
to check that \lstinline{arg[p]} does not change. The constructor stores a copy of \lstinline{arg[p]} and the returned base
contract compares the copy with the current value. 

\begin{lstlisting}[name=examples2]
var NotChangedCtor = Contract.Constructor(function(target) {
  var v = target[p];
  return Contract.Base(function (arg) {
    return (v === target[p]);
  });
});
\end{lstlisting}

\subsection{Implicit Assertions}

Naturally, each predicate is an implicit conjunction of different properties. 
Consider the base contracts \lstinline{|NonEmpty|} and \lstinline{|NotEmpty|}, 
both of which check the length of an array.

\begin{lstlisting}
var |NonEmpty| = Contract.Base(function (arg) {
  return (arg.length>0);
});
var |NotEmpty| = Contract.Base(function (arg) {
  return (arg.length!==0);
});
\end{lstlisting}

Described in detail, both contracts require that the given argument is an object 
\emph{and} the comparison of the arguments property \lstinline{length} and the 
value \lstinline{0} results in true. The comparison produces \lstinline{true}, \lstinline{false}, 
or \lstinline{undefined}, where  \lstinline{undefined} indicates that at least one operand is 
\lstinline{NaN}. But, the meaning of the contracts \lstinline{|NonEmpty|} and \lstinline{|NotEmpty|} 
is different. Both check that the arguments length is not zero. But, if the given 
argument does not have a length property or the property is, for example, a string value, the predicate 
in \lstinline{|NotEmpty|} holds whereas \lstinline{|NonEmpty|} fails. 

%% file: sm.body.syntax.tex
%
%

\section{JavaScript Core Calculus with Contracts}
\label{sec:formalization}

This section introduces $\lcon$, a call-by-value lambda calculus with
objects and object proxies that serves as a core calculus for JavaScript, inspired by previous work
\cite{GuhaSaftoiuKrishnamurthi2010,HeideggerBieniusaThiemann2012-popl,KeilThiemann2013-Proxy,CutsemMiller2013}.
It defines its syntax and describes its semantics informally.


\subsection{Core Syntax of $\lcon$}
\label{sec:syntax}

\begin{figure*}
  \vspace{-\baselineskip}
  \begin{displaymath}
    \begin{array}{ll@{~}r@{~}l}
      \Label{Expression} &\ni\ljExp,\ljExpf,\ljExpg
      &\bbc& \ljConst 
      \mid \ljVar 
      \mid \conE
      \mid \conA
      \mid \ljOp(\ljExp, \ljExpf)
      \mid \ljFunction
      \mid \ljExp(\ljExpf)
      \mid \ljNew\,\ljExp
      \mid \ljExp[\ljExpf]
      \mid \ljExp[\ljExpf]=\ljExpg 
      \mid \ljExp\ljAtl{}\ljExpf
      \\

      \\
      \Label{Contract Expressions} &\ni\conE,\conF &\bbc& \defBase\ljVar\ljExp
      \mid \defFun\conE\conF
      \mid \defDep\ljVar\conA
      \mid \defCap\conE\conF
      \mid \defCup\conE\conF
      \mid \defAnd\conE\conF
      \mid \defOr\conE\conF
      \mid \defNeg\conE
      \mid \conM\\

      \\
      \Label{Contract} &\ni\con,\conD &\bbc& \conI \mid \conQ 
      \mid \defCap\conI\con \mid \defCup\con\conD \mid \defAnd\con\conD
      \mid \defOr\conI\con\\

      \\
      \Label{Immediate} &\ni\conI,\conJ &\bbc& \defBase\ljVar\ljExp
      \mid \defNeg\conI\\

      \Label{Delayed} &\ni\conQ,\conR &\bbc& \defFun\con\conD
      \mid \defDep\ljVar\conA
      \mid \conM
      \mid \defCap\conQ\conR
      \mid \defOr\conQ\conR \mid \defNeg\conQ\\

      \\
      \Label{Abstraction} &\ni\conA &\bbc& \defAbs\ljVar\conE \\
      \Label{Mapping}& \ni\conM &\bbc& \emptyset \mid \defMap{\conM}{\ljConst}{\conE}\\

    \end{array}
  \end{displaymath}
  \caption{Syntax of $\lcon$.}
  \label{fig:syntax_lcon}
\end{figure*}

Figure~\ref{fig:syntax_lcon} defines the syntax of $\lcon$.
A $\lcon$ expression is either a constant, a variable, a contract expression,
an contract abstraction, an operation on primitive values, 
a lambda abstraction, an application, a creation of an empty object, 
a property read, a property assignment, or a contract assertion
that performs contract monitoring.
Variables $\ljVar$,$\ljVary$ are drawn from denumerable sets of symbols and
constants $\ljConst$ include JavaScript's primitive values like numbers, strings,
booleans, as well as \emph{undefined} and \emph{null}.

The assert expression, $\ljExp\ljAt\ljExpf$, is new to our calculus. It evaluates 
expression $\ljExpf$ to a contract $\C$ and attaches $\C$ to the value of 
expression $\ljExp$.


\subsection{Contract normalization}
\label{sec:normalization}

Contracts $\con,\conD$ are drawn from a set of contract expressions $\conE,\conF$ that also contain 
top-level intersections of contracts.
But, contract monitoring first normalizes contracts 
into a canonical form before it starts their enforcement.

Normalization factorizes a contract into its immediate
part and its delayed part. The immediate part consists of flat contracts (viz.\
nonterminal $\conI$ and $\conJ$) that are subject to intersection whereas the
delayed part is an intersection of function contracts (viz.\ nonterminals $\conQ$ and $\conR$). Unions
are pulled out of intersections by applying the distributive law suitably. Negations 
apply DeMorgan's Law until the negations reaches an immediate or delayed contract. 


\subsection{Contracts and Contract Assertion}
\label{sec:contracts}

A \emph{canonical contract} $\con$ is either an immediate contract $\conI$, a
delayed contract $\conQ$,
an intersection between two contracts $\defCap{\conI}{\con}$, or
an union $\defCup{\con}{\conD}$,
or a boolean combination of 
contracts: conjunction $\defAnd{\con}{\conD}$, disjunction
$\defOr{\conI}{\con}$, or negation $\defNot{\conI}$ or $\defNot{\conQ}$. 

Contract distinguish immediate contracts $\conI$ that may be checked 
right away when asserted to a value and delayed contracts $\conQ$ that need 
to be checked later on. 

The \emph{base contract} $\defBase\ljVar\ljExp$ is immediate. 
It consists of a function $\ljFunction$ that is interpreted as a predicate on $\ljVar$. 
Asserting base contract to a value applies the function to the value and
signals a violation unless the function returns a truthy
result. Otherwise, the assertion returns the value.

The evaluation of the predicate's body $\ljExp$ takes place in a sandbox.
The sandbox ensures noninterference with the actual program code
by stopping evaluation of the predicate if there is
an attempt to modify data that is visible outside the contract.%
\footnote{The sandbox cannot ensure termination of the predicate, of course.}

A delayed contract $\conQ$ is either a function contract $\defFun\con\conD$,
a dependent contract $\defDep\ljVar\conA$, an object contract $\conM$, 
an intersection of delayed contracts $\defCap\conQ\conR$, or a disjunction
of delayed contracts $\defOr\conQ\conR$.

The delayed contracts include intersections and disjunctions because each application 
of a function with such a contract has to check all parts of the contract.

A \emph{function contract} $\defFun\con\conD$ is built from a pair of contracts,
one for the domain ($\con$ is asserted to the argument) and one for the
range ($\conD$ is asserted to the result) of a function.  

A \emph{dependent contract} $\defDep\ljVar\conA$ is a kind of function contract.
Applying a function contracted to some value $\ljVal$ applies the abstraction 
$\conA$ to $\ljVal$ and asserts the resulting contract to
the result of applying the original function to $\ljVal$.

An \emph{object contract} $\conM$ is a mapping from property names to
contracts. The contract for property $\ljConst$ is expected to hold for
property $\ljConst$ of the contracted object. In practice, $\ljConst$
is a string.

A \emph{contract abstraction} $\defAbs\ljVar\C$ is not a contract by
itself, but abstracts the parameter $\ljVar$ that is
bound in contract $\C$. Dependent contracts and constructor applications
rely on abstractions to substitute run-time values for $\ljVar$ in $\C$.


\subsection{Constraints}
\label{sec:constraints}

Precise blaming of violators gives a useful feedbacks to the developers. 
However, signaling a violation depends on a combination of failures and successes 
in different contracts and is not necessarily the last failing contract.

\begin{figure*}[tp]
  \vspace{-\baselineskip}
  \begin{displaymath}
    \begin{array}{llrl}


      \Label{Identifier} &\ni~\cbIdent &\bbc& \cbBlame \mid \cbVar\\

      \Label{Constraints} &\ni~\cb 
      &\bbc& \defCb{\cbIdent}{\ljVal}
      \mid \defCb{\cbIdent}{\defFun{\cbVar}{\cbVar}}
      \mid \defCb{\cbIdent}{\defSet{\cbVar}}
      \mid \defCb{\cbIdent}{\defCap{\cbVar}{\cbVar}}
      \mid \defCb{\cbIdent}{\defCup{\cbVar}{\cbVar}}
      \mid \defCb{\cbIdent}{\defAnd{\cbVar}{\cbVar}}
      \mid \defCb{\cbIdent}{\defOr{\cbVar}{\cbVar}}
      \mid \defCb{\cbIdent}{\defNeg{\cbVar}}\\

      \Label{State} &\ni~\cbState &\bbc& \emptyset \mid \cbState\cup\cbState \mid \{\cb\}
    \end{array}
  \end{displaymath}
  \caption{Syntax of callbacks.}
  \label{fig:callback-syntax}
\end{figure*}

This connection is modeled by so-called \emph{constrains} $\cb$ 
(see Figure~\ref{fig:callback-syntax}). They are tied to a particular
contract assertion and link each contract to its next enclosing boolean
operation, at the top-level, its assertion. 

Constraints contain blame identifiers $\cbIdent$, where we distinguish blame
labels $\cbBlame$ that occur in source programs from internal blame variables $\cbVar$. Each blame
identifier is related to one specific contract occurrence in the program; there is at
least one identifier for each contract occurrence and there may be multiple identifiers for
delayed contracts (e.g., function contracts).

Each blame identifier $\cbIdent$ is associated with a record that defines the blame assignment for the contract
occurrence related to  $\cbIdent$. This record contains two fields, $\SubjectName$ and
$\ContextName$. The intuitive meaning of the fields is as follows. If the $\SubjectName$ field is false,
then the contract fails blaming the subject (i.e., the value to which the contract is asserted). If
the $\ContextName$ field is false, then the contract fails blaming the context (i.e., the user of
the value to which the contract is asserted).

A constraint is either the base constraint $\defCb{\cbIdent}{\ljVal}$ that signals the outcome 
$\ljVal$ of a base contract to blame identifier $\cbIdent$  or it chains the outcomes of 
the constituents of a contract to the outcome of the contract in $\cbIdent$.

The fields in the outcome range over $\cbLattice_4$, the lattice underlying Belnap's
four-valued logic.
Belnap~\cite{Belnap1977} introduced a logic intended to deal with incomplete or 
inconsistent information. The set $\cbLattice_4=\{\cbBot, \cbFalse, \cbTrue, \cbTop\}$ of 
truth values models a lattice illustrating accumulated knowledge about the truth 
of a proposition. Thus, a truth value may be considered as the set of classical truth values
$\{\cbTrueL, \cbFalseL\}$ that have been observed so far for a proposition.
For instance, contracts are counted ad $\cbBot$ before they are evaluated and $\cbTop$
signals potentially conflicting outcomes of repeated checking of the same contract.

The set $\cbLattice_4$ naturally forms a powerset lattice, the \emph{approximation
lattice}, with the subset ordering on the underlying sets, which we denote by
$\sqsubseteq$, and an ordering by truth, the \emph{logical lattice}.
We use the symbols $\cbNotOp$, $\cbAndOp$, and $\cbOrOp$ for
the connectives on $\cbLattice_4$; they are generalizations of the standard connectives on truth.
Meet and join under $\sqsubseteq$ are denoted by $\sqcap$ and $\sqcup$.

Boolean terms are standard, but their interpretation is in terms 
of $\cbLattice_4$. Similar to other logics, conjunction and disjunction 
have ``shortcuts'' that enable them to fail (respectively, succeed) 
even if one of the operands is still $\cbBot$.
More informations about boolean connectives in 
a bilattce-based logic can be found in~\cite{Majkic2005}.

Contract monitoring happens in the context of a \emph{constraint set} $\cbState$ which collects 
constraints during evaluation.

A \emph{solution} $\cbSolution$ of a constraint set $\cbState$ is a mapping from blame identiers to
records of elements of $\cbLattice_4$, such that all constraints are satisfied. Formally, we specify
the mapping by\footnote{We write $\Rangeof{\cbIdent}$ for the set ranged over by metavariable $\cbIdent$.}
\begin{displaymath}
  \cbSolution \in
  (\Rangeof{\cbIdent} \times \{\SubjectName, \ContextName\})
  \to \cbLattice_4
\end{displaymath}
and constraint satisfaction by a relation $\cbSolution \cbSatisfies \cbState$, which is specified in
Figure~\ref{fig:constraint-satisfaction}.
\begin{figure}[tp]
  \vspace{-\baselineskip}
  \begin{mathpar}
    \inferrule[\RuleCSEmpty]{%
    }{%
      \cbSolution \cbSatisfies \emptyset
    }

    \inferrule[\RuleCSUnion]{%
      \cbSolution \cbSatisfies \cbState_1\\
      \cbSolution \cbSatisfies \cbState_2
    }{%
      \cbSolution \cbSatisfies \cbState_1 \cup \cbState_2
    }

    \inferrule[\RuleCTFlat]{%
      \cbSolution (\subject\cbIdent) \sqsupseteq \cbMakeVal{\ljBer}\\
      \cbSolution (\context\cbIdent) \sqsupseteq \cbTrue
    }{%
      \cbSolution \cbSatisfies \cbIdent \update \ljBer
    }

    \inferrule[\RuleCTFunction]{%
      \cbSolution(\subject\cbIdent) \sqsupseteq 
      \cbSolution(\context{\cbVar_1} \cbAndOp (\subject{\cbVar_1} \cbImpOp
      \subject{\cbVar_2}))\\
      \cbSolution(\context\cbIdent) \sqsupseteq
      \cbSolution(\subject{\cbVar_1} \cbAndOp \context{\cbVar_2})
    }{%
      \cbSolution \cbSatisfies \cbIdent \update \defFC{\cbVar_1}{\cbVar_2}
    }

    \inferrule[\RuleCTSet]{%
      \cbSolution(\subject\cbIdent) \sqsupseteq 
      \cbTrue\\
      \cbSolution(\context\cbIdent) \sqsupseteq
      \cbSolution(\subject{\cbVar})
    }{%
      \cbSolution \cbSatisfies \cbIdent \update \defSet{\cbVar}{}
    }

    \inferrule[\RuleCTIntersection]{%
      \cbSolution(\subject\cbIdent) \sqsupseteq \cbSolution(\subject{\cbVar_1} \cbAndOp \subject{\cbVar_2}) \\
      \cbSolution(\context\cbIdent) \sqsupseteq \cbSolution(\context{\cbVar_1} \cbOrOp \context{\cbVar_2})
    }{%
      \cbSolution \cbSatisfies \cbIdent \update \defCap{\cbVar_1}{\cbVar_2}
    }

    \inferrule[\RuleCTUnion]{%
      \cbSolution(\subject\cbIdent) \sqsupseteq \cbSolution(\subject{\cbVar_1} \cbOrOp \subject{\cbVar_2})\\
      \cbSolution(\context\cbIdent) \sqsupseteq \cbSolution(\context{\cbVar_1} \cbAndOp \context{\cbVar_2})
    }{%
      \cbSolution \cbSatisfies \cbIdent \update \defCup{\cbVar_1}{\cbVar_2}
    }

    \inferrule[\RuleCTAnd]{%
      \cbSolution(\subject\cbIdent) \sqsupseteq \cbSolution(
      (\context{\cbVar_1} \cbAndOp \context{\cbVar_2}) \cbImpOp
      (\subject{\cbVar_1} \cbAndOp \subject{\cbVar_2})
      )\\
      \cbSolution(\context\cbIdent) \sqsupseteq \cbSolution(\context{\cbVar_1} \cbAndOp \context{\cbVar_2})
    }{%
      \cbSolution \cbSatisfies \cbIdent \update \defAnd{\cbVar_1}{\cbVar_2}
    }

    \inferrule[\RuleCTOr]{%
      \cbSolution(\subject\cbIdent) \sqsupseteq \cbSolution(
      (\context{\cbVar_1} \cbOrOp \context{\cbVar_2}) \cbImpOp
      ((\context{\cbVar_1} \cbAndOp \subject{\cbVar_1}) \cbOrOp
      (\context{\cbVar_2} \cbAndOp \subject{\cbVar_2}))
      )\\
      \cbSolution(\context\cbIdent) \sqsupseteq \cbSolution(\context{\cbVar_1} \cbOrOp \context{\cbVar_2})
    }{%
      \cbSolution \cbSatisfies \cbIdent \update \defOr{\cbVar_1}{\cbVar_2}
    }

    \inferrule[\RuleCTNeg]{%
      \cbSolution(\subject\cbIdent) \sqsupseteq \cbSolution(\subject{\cbVar_1} \cbImpOp \cbNotOp \context{\cbVar_1})\\
      \cbSolution(\context\cbIdent) \sqsupseteq \cbTrue
    }{%
      \cbSolution \cbSatisfies \cbIdent \update \defNeg{\cbVar}
    }

  \end{mathpar}
  \caption{Constraint satisfaction.}
  \label{fig:constraint-satisfaction}
\end{figure}

In the premisses, the rules apply a constraint mapping $\cbSolution$ to boolean expressions over constraint
variables. This application stands for the obvious homomorphic extension of the mapping.

Rule \RefTirName{\RuleCSEmpty} states that 
every mapping satisfies the empty set of constraints.
Rule \RefTirName{\RuleCSUnion} states that the union of constraints corresponds to the intersection of
sets of solutions. 

The rule \RefTirName{\RuleCTFlat} treats the constraint generated for the outcome $\ljBer$ of the
predicate of a flat contract. The function $\cbMakeVal{\cdot} : \Rangeof\ljBer \to \cbLattice_4$
translates outcomes to truth values. It corresponds to JavaScript's idea of \emph{truthy} and
\emph{falsey}. It is defined by 
\begin{align*}
  \cbMakeVal{\ljBer} \df \begin{cases}
    \cbFalse, & \ljBer\in\{\ljExn,\ljFalse,\ljUndefined,\ljNull,\ljNaN,\ljEmyStr\}\\
    \cbTrue, & \text{otherwise}
  \end{cases}
\end{align*}
where $\ljExn$ symbolises an exception, that counts as false, too.

The rule \RefTirName{\RuleCTFunction} determines the blame assignment for a function contract $\cbIdent$
from the blame assignment for the argument and result contracts, which are available through
$\cbVar_1$ and $\cbVar_2$. A function does not fulfill contract $\cbIdent$ if 
it does not fulfill its obligations towards its argument $\context{\cbVar_1}$ \textbf{or}
if the argument fulfills its contract, but the result does not fulfills its contract.
The first part arises for higher-order functions, which may pass illegal arguments to their
function-arguments. The second part corresponds to partial correctness of the function with 
respect to its contract. 

A function's context (caller) fulfills the contract if
it passes an argument that fulfills its contract $\subject{\cbVar_1}$ \textbf{and}
it uses the function result according to its contract $\context{\cbVar_2}$.
The second part becomes non-trivial with functions that return functions.

Rule \RefTirName{\RuleCTSet} determines the blame assignment for an object contract at $\cbIdent$
from the blame assignment for the property contract at $\cbVar$.
An object (subject) does not fulfill an object contract if it returns a value that does not fulfill 
the contract. An object's context (caller) does not fulfills the contract if it does not uses 
the objects return according to its contract \textbf{or} is assigns an illegal value to 
a contracted property.

The outcome of read access on contracted properties is directly related
to $\cbIdent$ and do not need a special constraint.
For write access, it is up to the context to assign a value 
according to the contract. Constraint $\defCb{\cbIdent}{\defSet{\cbVar}}$ 
gives the responsibilities of $\subject{\cbVar}$ to the context $\context\cbIdent$. 

The rule \RefTirName{\RuleCTIntersection} determines the blame assignment
for an intersection contract at $\cbIdent$ from its constituents at
$\cbVar_1$ and $\cbVar_2$. 
A subject fulfills an intersection contract
if it fulfills both constituent contracts: $\subject{\cbVar_1}
\cbAndOp \subject{\cbVar_2}$. A context, however, only needs to
fulfill one of the constituent contracts:  $\context{\cbVar_1}
\cbOrOp \context{\cbVar_2}$.

Dually to the intersection rule, the rule \RefTirName{\RuleCTUnion}
determines the blame assignment 
for a union contract at $\cbIdent$ from its constituents at
$\cbVar_1$ and $\cbVar_2$. A subject fulfills a union contract
if it fulfills one of the constituent contracts: $\subject{\cbVar_1}
\cbOrOp \subject{\cbVar_2}$. A context, however, needs to
fulfill both constituent contracts:  $\context{\cbVar_1}
\cbAndOp \context{\cbVar_2}$, because it does not known which contract
is fulfilled by the subject. 

The rules \RefTirName{\RuleCTAnd}, \RefTirName{\RuleCTOr}, \RefTirName{\RuleCTNeg} show blame assignment
for the classical boolean connectives.
A subject fulfills a conjunction contract at $\cbIdent$ if it fulfills both contracts at
$\subject{\cbVar_1}$ and $\subject{\cbVar_2}$ in case the context fulfills its obligations.
The context also needs to fulfill both contracts: $\context{\cbVar_1}$ and $\context{\cbVar_2}$.

In the same way, a subject fulfills a disjunction contract at $\cbIdent$ if it fulfills at least one side 
of the disjunction. The context might choose to fulfill at $\context{\cbVar_1}$ or $\context{\cbVar_2}$.

For negations, a subject fulfills a negation contract at $\cbIdent$ if it does not fulfill 
the contract at $\subject{\cbVar}$ in case the context fulfills its obligation. 
The context does always fulfill its obligation.

Inspection of constraint satisfaction and monotonicity of the logical operations on $\cbLattice_4$
shows that every finite set of constraints corresponds to a monotone mapping on a complete lattice
of finite height. By the Knaster-Tarski theorem, each constraint set $\cbState$ has a least solution
\begin{displaymath}
  \Sem{\cbState}  \in
  (\Rangeof{\cbIdent} \times \{\SubjectName, \ContextName\})
  \to \cbLattice_4
\end{displaymath}
such that $\Sem{\cbState} \cbSatisfies \cbState$ and $\Sem{\cbState} \sqsubseteq \cbSolution$ for all
$\cbSolution \cbSatisfies \cbState$, which can be computed as the
least fixpoint of the monotone mapping. Hence, we define the semantics of a
constraint set $\Sem{\cbState}$ as this least fixpoint.

To determine whether a constraint set $\cbState$ is a blame state (i.e., whether it should signal a contract
violation), we need to check whether the semantics $\Sem\cbState$ maps any source level blame label
$\cbBlame$ to false.
\begin{definition}[Blame State]
  $\cbState$ is a \emph{blame state} iff\newline
  $\Sem{\cbState}^{-1} (\{\cbFalse, \cbTop\}) \cap (\Rangeof{\cbBlame} \times 
  \{\SubjectName, \ContextName\}) \ne \emptyset $.
\end{definition}
\begin{lemma}\label{thm:blamestate}
  The \emph{blame state} predicate is a monotone mapping from $(\Rangeof{\cbIdent} \times
  \{\SubjectName, \ContextName\}) \to \cbLattice_4$ to $\{ \ljFalse, 
  \ljTrue \}$ ordered by $\ljFalse < \ljTrue$.
\end{lemma}

%% file: sm.body.monitoring.tex
%
%
\section{Contract Monitoring}
\label{sec:monitoring}

This section presents the formal semantics of $\lcon$ including
evaluation and contract enforcement. The formalization employs pretty-big-step semantics
\cite{Chargueraud2013} to model side effects and the 
signaling of violations while keeping the 
number of evaluation rules manageable. 


\subsection{Semantic Domains}
\label{sec:semantic_domains}

\begin{figure}[tp]
  \vspace{-\baselineskip}
  \begin{displaymath}
    \begin{array}{llrl}  
      \Label{Value} &\ni~\ljVal,\ljValu,\ljValw &\bbc& \ljConst \mid \ljLocation \mid (\ljSEnv,\C) \mid (\ljSEnv,\conA)\\

      \\
      \Label{Closure} &\ni~\ljClosure &\bbc& \ljNoClosure \mid (\ljEnv,\ljFunction)\\
      \Label{Dictionary} &\ni~\ljDic &\bbc& \emptyset \mid \ljDic[\ljConst\mapsto\ljVal]\\
      \Label{Proxy Handler} &\ni~\ljHandler &\bbc& (\ljSEnv) \mid (\ljSEnv, \cbIdent, \conQ)\\
      \Label{Object} &\ni~\ljObj &\bbc& (\ljDic, \ljClosure, \ljVal) \mid (\ljLocation, \ljHandler)\\

      \\
      \Label{Environment} &\ni~\ljEnv &\bbc& \emptyset \mid \ljEnv[\ljVar\mapsto\ljVal]\\
      \Label{Store} &\ni~\ljStore &\bbc& \emptyset \mid \ljStore[\ljLocation\mapsto\ljObj]\\

    \end{array}
  \end{displaymath}
  \vspace{-\baselineskip}
  \caption{Semantic domains of $\lcon$.}
  \label{fig:domains_lcon}
\end{figure}

Figure~\ref{fig:domains_lcon} defines the semantic domains of $\lcon$.

Its main component is a store that maps a location $\ljLocation$ to an object $\ljObj$, 
which is either a native object (non-proxy object) represented by a triple consisting of a 
dictionary $\ljDic$, a potential function closure $\ljClosure$, and a value $\ljVal$ acting 
as prototype or a proxy object. A dictionary $\ljDic$ models the properties of an object.
It maps a constant $\ljConst$ to a value $\ljVal$. An object may be a function in which case 
its closure consists of a lambda expression $\ljFunction$ and an environment $\ljEnv$ that 
binds the free variables. It maps a variable $\ljVar$ to a value $\ljVal$. A non-function object 
is indicated by $\ljNoClosure$ in this place. 

A proxy object is a single location controlled by a proxy handler $\ljHandler$ that mediates the access 
to the target location. For simplification, we represent handler objects by there meta-data. So, a 
handler is either sandbox handler that enforces write-protection (viz. by an single 
\emph{secure} environment $\ljSEnv$) or a contracted object (viz. by an environment $\ljSEnv$, a callback 
identifier $\cbIdent$, and a delayed contract $\conQ$). 

The syntax do not make proxies available to the user, but offers an internal method to wrap objects.

A value $\ljVal$ is either a constant $\ljConst$, a location $\ljLocation$, or a \emph{contract value}, which is 
either a contract $\con$ or a contract abstraction $\conA$ in combination with an sandbox environment.

For clarity, we write $\ljSVal$,$\ljSValu$,$\ljSValw$ for wrapped values that are imported into a 
sandbox and $\ljSEnv$ for a sandbox environment that only contains wrapped values.


\subsection{Evaluation of $\lcon$}
\label{sec:evaluationj}

\begin{figure}[tp]
  \vspace{-\baselineskip}
  \begin{displaymath}
    \begin{array}{llrl}
      \Label{Behavior} &\ni~\ljBer &\bbc& \ljVal \mid \ljExn \\
      \Label{Term} &\ni~\ljTerm &\bbc& \ljExp \mid \ljOp(\ljBer,\ljExp) \mid \ljOp(\ljBer,\ljBer) \mid \ljBer(\ljExp) \mid \ljBer(\ljBer)
      \mid \ljNew~\ljBer \mid \ljBer[\ljExp] \mid \ljBer[\ljBer] \mid \ljBer[\ljExp]\!=\!\ljExp \mid \ljBer[\ljBer]\!=\!\ljExp\\ 
      &&\mid& \ljBer[\ljBer]\!=\!\ljBer \mid \ljBer\ljAtb{}\ljExp \mid \ljBer\ljAtb{}\ljBer \mid \ljVal\ljAtb{}\con
      \mid \conA(\ljBer) \mid \ljWrap(\ljExp) \mid \ljWrap(\ljBer)
    \end{array}
  \end{displaymath}
  \vspace{-\baselineskip}
  \caption{Intermediate terms of $\lcon$.}
  \label{fig:terms_lcon}
\end{figure}

A pretty-big-step semantics introduces intermediate terms to model partially evaluated expressions
(Figure~\ref{fig:terms_lcon}). 
An intermediate term is thus an expression where zero or more top-level subexpressions are replaced by
their outcomes, which we call \emph{behaviors}. 
A behavior $\ljBer$ is either a value $\ljVal$ or an exception $\ljExn$, which may be associated
with a contract or a sandbox violation.
Terms are constructed with a specific evaluation order in mind so that the exception handling rules
that propagate exceptions to the top-level are easy to state.

The evaluation judgment is similar to a standard big-step evaluation judgment except that its input
ranges over intermediate terms and its output is a behavior: It states that evaluation of 
term $\ljTerm$ with constraint set $\cbState$, initial store $\ljStore$, and environment $\ljEnv$ 
results in a new constraint set $\cbState'$, final heap $\ljStore'$, and behavior $\ljBer$.
\begin{displaymath}
  \ljEnv \entails \langle \cbState,\ljStore,\ljTerm \rangle \eval \langle \cbState,\ljStore',\ljBer \rangle
\end{displaymath}

\begin{figure*}[tp]
  \centering
  \begin{mathpar}
    \inferrule [\RuleLjConst]
    {%
    }
    {%
      \ljEnv \entails \cbState,\ljStore, \ljConst \eval \cbState,\ljStore, \ljConst
    }\and
    \inferrule [\RuleLjVar]
    {%
    }
    {%
      \ljEnv \entails \cbState,\ljStore,\ljVar \eval \cbState,\ljStore,\ljEnv(\ljVar)
    }\and
    \inferrule [\RuleLjOpE]
    {%
      \ljEnv \entails \cbState,\ljStore,\ljExp \eval \cbState',\ljStore',\ljBer\\\\
      \ljEnv \entails \cbState',\ljStore',\ljOp(\ljBer, \ljExpf) \eval \cbState',\ljStore'',\ljBer'
    }
    {%
      \ljEnv \entails \cbState,\ljStore,\ljOp(\ljExp, \ljExpf) \eval \cbState',\ljStore'',\ljBer'
    }\and
    \inferrule [\RuleLjOpF]
    {%
      \ljEnv \entails \cbState,\ljStore,\ljExpf \eval \cbState',\ljStore',\ljBer\\\\
      \ljEnv \entailsOP \cbState',\ljStore',\ljOp(\ljVal, \ljBer) \eval \cbState',\ljStore',\ljBer'
    }
    {%
      \ljEnv \entails \cbState,\ljStore,\ljOp(\ljVal, \ljExpf) \eval \cbState',\ljStore',\ljBer'
    }\and
    \inferrule [\RuleLjOp]
    {%
      \ljValw=\ljOp(\ljVal, \ljValu)
    }
    {%
      \ljEnv\entailsOP\cbState,\ljStore,\ljOp(\ljVal,\ljValu) \eval \cbState,\ljStore,\ljValw
    }\and
    \inferrule [\RuleLjAbs]
    {%
      \ljLocation \notin \dom{\ljStore}\\
      \ljStore'=\ljStore[\ljLocation\mapsto(\emptyset,(\ljEnv, \ljFunction),\ljNull)]
    }
    {%
      \ljEnv \entails \cbState,\ljStore,\ljFunction \eval \cbState,\ljStore',\ljLocation
    }\and
    \inferrule [\RuleLjAppE]
    {%
      \ljEnv \entails \cbState,\ljStore,\ljExp \eval \cbState',\ljStore',\ljBer\\\\
      \ljEnv \entails \cbState',\ljStore',\ljBer(\ljExpf) \eval \cbState'',\ljStore'',\ljBer'
    }
    {%
      \ljEnv \entails \cbState,\ljStore,\ljExp(\ljExpf) \eval \cbState'',\ljStore'',\ljBer'
    }\and
    \inferrule [\RuleLjAppF]
    {%
      \ljEnv \entails \cbState,\ljStore,\ljExpf \eval \cbState',\ljStore',\ljBer\\\\
      \ljEnv \entailsFA \cbState',\ljStore',\ljLocation(\ljBer) \eval \cbState'',\ljStore'',\ljBer'
    }
    {%
      \ljEnv \entails \cbState,\ljStore,\ljLocation(\ljExpf) \eval \cbState'',\ljStore'',\ljBer'
    }\and
    \inferrule [\RuleLjApp]
    {%
      ( \ljDic, (\ljEnv', \ljFunction), \ljVal) = \ljStore(\ljLocation) \\
      \ljEnv'[\ljVar \mapsto \ljVal] \entails \cbState,\ljStore,\ljExp \eval \cbState',\ljStore',\ljBer
    }
    {%
      \ljEnv \entailsFA \cbState,\ljStore,\ljLocation(\ljVal) \eval \cbState',\ljStore',\ljBer
    }\and
    \inferrule [\RuleLjNewE]
    {%
      \ljEnv \entails \cbState,\ljStore,\ljExp \eval \cbState',\ljStore',\ljBer\\\\
      \ljEnv \entails \cbState',\ljStore',\ljNew~\ljBer \eval \cbState'',\ljStore'',\ljBer'
    }
    {%
      \ljEnv \entails \cbState,\ljStore,\ljNew~\ljExp \eval \cbState'',\ljStore'',\ljBer'
    }\and
    \inferrule [\RuleLjNew]
    {%
      \ljLocation \notin \dom{\ljStore}\\
      \ljStore'=\ljStore[\ljLocation\mapsto(\emptyset,\ljNoClosure,\ljVal)]
    }
    {%
      \ljEnv \entails \cbState,\ljStore,\ljNew~\ljVal \eval \cbState,\ljStore',\ljLocation
    }\and
    \inferrule [\RuleLjGetE]
    {%
      \ljEnv \entails \cbState,\ljStore,\ljExp \eval \cbState',\ljStore',\ljBer\\\\
      \ljEnv \entails \cbState',\ljStore',\ljBer[\ljExpf] \eval \cbState'',\ljStore'',\ljBer'
    }
    {%
      \ljEnv \entails \cbState,\ljStore,\ljExp[\ljExpf] \eval \cbState'',\ljStore'',\ljBer'
    }\and
    \inferrule [\RuleLjGetF]
    {%
      \ljEnv \entails \cbState,\ljStore,\ljExpf \eval \cbState',\ljStore',\ljBer\\\\
      \ljEnv \entailsPR \cbState',\ljStore',\ljLocation[\ljBer] \eval \cbState'',\ljStore'',\ljBer'
    }
    {%
      \ljEnv \entails \cbState,\ljStore,\ljLocation[\ljExpf] \eval \cbState'',\ljStore'',\ljBer
    }\and
    \inferrule [\RuleLjGet]
    {%
      (\ljDic,\ljClosure,\ljVal) = \ljStore(\ljLocation)\\
      \ljConst\in\dom\ljDic
    }
    {%
      \ljEnv \entailsPR \cbState,\ljStore,\ljLocation[\ljConst] \eval \cbState,\ljStore,\ljDic(\ljConst)
    }\and
    \inferrule [\RuleLjGetPrototype]
    {%
      (\ljDic,\ljClosure,\ljLocation) = \ljStore(\ljLocation)\\
      \ljConst\notin\dom\ljDic\\\\
      \ljEnv \entailsPR \cbState,\ljStore,\ljLocation'[\ljConst] \eval \cbState',\ljStore',\ljBer
    }
    {%
      \ljEnv \entailsPR \cbState,\ljStore,\ljLocation[\ljConst] \eval \cbState',\ljStore',\ljBer
    }\and
    \inferrule [\RuleLjGetUndefined]
    {%
      (\ljDic,\ljClosure,\ljConst) = \ljStore(\ljLocation)\\
      \ljConst\notin\dom\ljDic 
    }
    {%
      \ljEnv \entailsPR \cbState,\ljStore,\ljLocation[\ljConst] \eval \cbState,\ljStore,\ljUndefined
    }\and
    \inferrule [\RuleGetNoContract]
    {%
      (\ljLocation',\ljSEnv,\cbIdent,\conM) = \ljStore(\ljLocation)\\
      \ljConst\not\in\dom{\OC}\\\\
      \ljEnv \entailsPR \cbState,\ljStore,\ljLocation'[\ljConst] \eval \cbState',\ljStore',\ljBer
    }
    {%
      \ljEnv \entailsPR \cbState,\ljStore,\ljLocation[\ljConst] \eval \cbState',\ljStore',\ljBer
    }
  \end{mathpar}
  \caption{Inference rules for $\lcon$.}
  \label{fig:lcon}
\end{figure*}

\begin{figure*}[tp]
  \centering
  \begin{mathpar}
    \inferrule [\RuleLjPutE]
    {%
      \ljEnv \entails \cbState,\ljStore,\ljExp \eval \cbState',\ljStore',\ljBer\\\\
      \ljEnv \entails \cbState',\ljStore',\ljBer[\ljExpf]\!=\!\ljExpg \eval \cbState'',\ljStore'',\ljBer'
    }
    {%
      \ljEnv \entails \cbState,\ljStore,\ljExp[\ljExpf]\!=\!\ljExpg \eval \cbState'',\ljStore'',\ljBer'
    }\and
    \inferrule [\RuleLjPutF]
    {%
      \ljEnv \entails \cbState,\ljStore,\ljExpf \eval \cbState',\ljStore',\ljBer\\\\
      \ljEnv \entails \cbState',\ljStore',\ljLocation[\ljBer]\!=\!\ljExpg \eval \cbState'',\ljStore'',\ljBer'
    }
    {%
      \ljEnv \entails \cbState,\ljStore,\ljLocation[\ljExpf]\!=\!\ljExpg \eval \cbState'',\ljStore'',\ljBer'
    }\and
    \inferrule [\RuleLjPutG]
    {%
      \ljEnv \entails \cbState,\ljStore,\ljExpg \eval \cbState',\ljStore',\ljBer\\\\
      \ljEnv \entailsPA \cbState',\ljStore',\ljLocation[\ljConst]\!=\!\ljBer \eval \cbState'',\ljStore'',\ljBer'
    }
    {%
      \ljEnv \entails \cbState,\ljStore,\ljLocation[\ljConst]\!=\!\ljExpg \eval \cbState'',\ljStore'',\ljBer'
    }\and
    \inferrule [\RuleLjPut]
    {%
      (\ljDic,\ljClosure,\ljVal) = \ljStore(\ljLocation)\\
      \ljStore'=\ljStore[\ljLocation\mapsto(\ljDic[\ljConst\mapsto\ljVal],\ljClosure,\ljVal)]
    }
    {%
      \ljEnv \entailsPA \cbState,\ljStore,\ljLocation[\ljConst]\!=\!\ljVal \eval \cbState,\ljStore',\ljVal
    }\and
     \inferrule [\RulePutNoContract]
    {%
      (\ljLocation',\ljSEnv,\cbIdent,\conM) = \ljStore(\ljLocation)\\
      \ljConst\not\in\dom{\conM}\\\\
      \ljEnv \entailsPA \cbState,\ljStore,\ljLocation'[\ljConst]\!=\!\ljVal \eval \cbState',\ljStore',\ljVal
    }
    {%
      \ljEnv \entailsPA \cbState,\ljStore,\ljLocation[\ljConst]\!=\!\ljVal \eval \cbState',\ljStore',\ljVal
    }\and
    \inferrule [\RuleLjAssertE]
    {%
      \ljEnv \entails \cbState,\ljStore,\ljExp \eval \cbState',\ljStore',\ljBer\\\\
      \ljEnv \entails \cbState',\ljStore',\ljBer\ljAt^{\cbBlame}\ljExpf \eval \cbState'',\ljStore'',\ljBer'
    }
    {%
      \ljEnv \entails \cbState,\ljStore,\ljExp\ljAt^{\cbBlame}\ljExpf \eval \cbState'',\ljStore'',\ljBer'
    }\and
    \inferrule [\RuleLjAssertF]
    {%
      \ljEnv \entails \cbState,\ljStore,\ljExp \eval \cbState',\ljStore',\ljBer\\\\
      \ljSEnvEmpty \entailsC \cbState,\ljStore',\ljVal\ljAt^{\cbBlame}\ljBer \eval \cbState'',\ljStore'',\ljBer'
    }
    {%
      \ljEnv \entails \cbState,\ljStore,\ljVal\ljAt^{\cbBlame}\ljExp \eval \cbState'',\ljStore'',\ljBer'
    }\and
    \inferrule [\RuleLjConstructF]
    {%
      \ljEnv \entails \cbState,\ljStore,\ljExpf \eval \cbState', \ljStore',\ljBer\\\\
      \ljEnv \entails \cbState',\ljStore',\defApp{(\ljSEnv,\defAbs{\ljVar}{\ljExp})}{\ljBer} \eval \cbState'', \ljStore'',\ljBer'\\
    }
    {%
      \ljEnv \entailsC \cbState,\ljStore,\defApp{(\ljSEnv,\defAbs{\ljVar}{\ljExp})}{\ljExpf} \eval \cbState'', \ljStore'',\ljBer'
    }\and
    \inferrule [\RuleLjWrapE]
    {%
      \ljEnv \entails \cbState,\ljStore,\ljExp \eval \cbState', \ljStore',\ljBer\\\\
      \ljEnv \entails \cbState',\ljStore',\ljWrap(\ljBer) \eval \cbState'', \ljStore'',\ljBer'\\
    }
    {%
      \ljEnv \entailsC \cbState,\ljStore,\ljWrap(\ljExp) \eval \cbState'', \ljStore'',\ljBer'
    }
  \end{mathpar}
  \caption{Inference rules for $\lcon$ (cont'd).}
  \label{fig:lcon2}
\end{figure*}

\begin{figure*}[tp]
  \centering
  \begin{mathpar}
    \inferrule[\RuleLjOpE\RuleError]
    {%
    }
    {%
      \ljEnv\entails\cbState,\ljStore,\ljOp(\ljExn, \ljExpf) \eval\cbState,\ljStore,\ljExn%
    }\and
    \inferrule[\RuleLjOpF\RuleError]
    {%
    }
    {%
      \ljEnv\entails\cbState,\ljStore,\ljOp(\ljVal, \ljExn) \eval\cbState,\ljStore,\ljExn%
    }\and 
    \inferrule [\RuleLjAppE\RuleError]
    {%
    }
    {%
      \ljEnv \entails \cbState,\ljStore,\ljExn(\ljExpf) \eval \cbState,\ljStore,\ljExn
    }\and
    \inferrule [\RuleLjAppF\RuleError]
    {%
    }
    {%
      \ljEnv \entailsFA \cbState,\ljStore,\ljLocation(\ljExn) \eval \cbState,\ljStore,\ljExn
    }\and
    \inferrule [\RuleLjNewE\RuleError]
    {%
    }
    {%
      \ljEnv \entails \cbState,\ljStore,\ljNew~\ljExn \eval \cbState,\ljStore,\ljExn
    }\and
    \inferrule [\RuleLjGetE\RuleError]
    {%
    }
    {%
      \ljEnv \entails \cbState,\ljStore,\ljExn[\ljExpf] \eval \cbState,\ljStore,\ljExn
    }\and
    \inferrule [\RuleLjGetF\RuleError]
    {%
    }
    {%
      \ljEnv \entailsPR \cbState,\ljStore,\ljLocation[\ljExn] \eval \cbState,\ljStore,\ljExn
    }\and
    \inferrule [\RuleLjPutE\RuleError]
    {%
    }
    {%
      \ljEnv \entails \cbState,\ljStore,\ljExn[\ljExpf]\!=\!\ljExpg \eval \cbState,\ljStore,\ljExn
    }\and
    \inferrule [\RuleLjPutF\RuleError]
    {%
    }
    {%
      \ljEnv \entails \cbState,\ljStore, \ljLocation[\ljExn]=\ljExpg \eval \cbState,\ljStore,\ljExn
    }\and
    \inferrule [\RuleLjPutG\RuleError]
    {%
    }
    {%
      \ljEnv \entailsPA \cbState,\ljStore, \ljLocation[\ljConst]=\ljExn \eval \cbState,\ljStore,\ljExn
    }\and
    \and
    \inferrule [\RuleLjAssertE\RuleError]
    {%
    }
    {%
      \ljEnv \entailsC \cbState,\ljStore,\ljExn\ljAt\ljExpf \eval \cbState,\ljStore,\ljExn
    }\and
    \inferrule [\RuleLjAssertF\RuleError]
    {%
    }
    {%
      \ljEnv \entailsC \cbState,\ljStore,\ljVal\ljAt\ljExn \eval \cbState,\ljStore,\ljExn
    }\and
    \inferrule [\RuleLjWrapE\RuleError] 
    {%
    }
    {%
      \ljSEnv \entailsWP \cbState,\ljStore,\ljWrap(\ljExn) \eval \cbState,\ljStore,\ljExn
    }\and
    \inferrule [\RuleLjConstructF\RuleError]
    {%
    }
    {%
      \ljSEnv \entailsC \cbState,\ljStore,\defApp{(\ljSEnv,\defAbs{\ljVar}{\ljExp})}{\ljExn}\ljAt\ljVal \evalc{\cb} \cbState,\ljStore,\ljExn
    }
  \end{mathpar}
  \caption{Inference rules for exception handling.}
  \label{fig:lcon_error}
\end{figure*}

Figure~\ref{fig:lcon} and~\ref{fig:lcon2} defines the standard evaluation rules for expressions $\ljExp$ in $\lcon$. 
The inference rules for expressions are mostly standard. Each rule for a composite expression evaluates 
exactly one subexpression and then recursively invokes the evaluation judgment to continue. 
Once all subexpressions are evaluated, the respective rule performs the desired operation.

The corresponding straightforward error propagation rules are disjoint to the remaining rules because 
they fire only if an intermediate term contains an exception. 
The rules set in Figure~\ref{fig:lcon_error} defines error handling.

\subsubsection{Contract Construction}

\begin{figure}[tp]
  \centering
  \begin{mathpar}
    \inferrule [\RuleLjContractFresh]
    {}
    {%
      \ljEnv \entails \cbState,\ljStore,\con \eval \cbState,\ljStore,(\emptyset,\con)
    }\and
    \inferrule [\RuleLjContractSbx]
    {}
    {%
      \ljSEnv \entails \cbState,\ljStore,\con \eval \cbState,\ljStore,(\ljSEnv,\con)
    }\and
    \inferrule [\RuleLjConstructorFresh]
    {}
    {%
      \ljEnv \entails \cbState,\ljStore,\conA \eval \cbState,\ljStore,(\emptyset,\conA)
    }\and
    \inferrule [\RuleLjConstructorSbx]
    {}
    {%
      \ljSEnv \entails \cbState,\ljStore,\conA \eval \cbState,\ljStore,(\ljSEnv,\conA)
    }\and
    \inferrule [\RuleLjConstruct]
    {%
      \ljEnv \entails \cbState,\ljStore,\ljWrap(\ljValw) \eval \cbState', \ljStore',\ljSValw\\
      \ljSEnv[\ljVar\mapsto\ljSValw] \entails \cbState',\ljStore',\con \eval \cbState',\ljStore',\ljBer\\
    }
    {%
      \ljEnv \entailsC \cbState,\ljStore,\defApp{(\ljSEnv,\defAbs{\ljVar}{\con})}{\ljValw} \eval \cbState', \ljStore',\ljBer
    }
  \end{mathpar}
  \caption{Contract Construction rules for $\lcon$.}
  \label{fig:lcon_ctor}
\end{figure}

Contract monitoring happens in the context of a secure sandbox environment to preserve 
noninterference.
So a contract definition (contract abstraction) will evaluate to a contract closure containing 
the contract (the abstraction) together with an empty environment 
or together with an sandbox environment $\ljSEnv$
when defining the contract (the abstraction) inside of a sandbox environment.
Figure~\ref{fig:lcon_ctor} contains its inference rules.

Constructor application (Rule \RefTirName{\RuleLjConstruct}) starts 
after the first expression evaluates to a contract closure and the second expression 
evaluates to a value. It wraps the given value and evaluates the contract $\con$ 
in the sandbox environment after binding the sandboxed value $\ljSValw$.

\subsubsection{Contract Assertion}

\begin{figure}[tp]
  \centering
  \begin{mathpar}
    \inferrule [\RuleLjAssert]
    {%
      \ljSEnv \entails \cbState,\ljStore,\ljVal\ljAtb{}\con \eval \cbState',\ljStore',\ljBer
    }
    {%
      \ljEnv \entails \cbState,\ljStore,\ljVal\ljAtb{}(\ljSEnv,\con) \eval \cbState',\ljStore',\ljBer
    }
  \end{mathpar}
  \hrule
  \begin{mathpar}
    \inferrule [\RuleAssertBase]
    {%
      \ljSEnv \entailsC \cbState,\ljStore,(\defBase{\ljVar}{\ljExp})(\ljWrap(\ljVal)) \eval \cbState',\ljStore',\ljBer\\\\
      \ljSEnv \entailsC \cbState'\cup\{\defCb{\cbIdent}{\cbMakeVal{\ljBer}}\},\ljStore',\ljVal\ljAtb{}\ljBer 
      \eval \cbState'',\ljStore'',\ljBer'
    }
    {%
      \ljSEnv \entailsC \cbState,\ljStore,\ljVal\ljAtb{}\defBase{\ljVar}{\ljExp} 
      \eval \cbState'',\ljStore'',\ljBer'
    }\and
    \inferrule [\RuleAssertDelayed]
    {%
      \ljLocation' \notin \dom{\ljStore}\\ 
      \ljStore' = \ljStore[\ljLocation'\mapsto (\ljLocation,\ljSEnv,\cbIdent,\conQ)]
    }
    {%
      \ljSEnv \entailsC \cbState,\ljStore,\ljLocation\ljAtb{}\conQ \eval \cbState,\ljStore',\ljLocation'
    }\and
    \inferrule [\RuleAssertIntersection]
    {%
      \cbVar_1,\cbVar_2\not\in\dom{\cbState}\\
      \ljSEnv \entailsC  \cbState\cup\{\defCb{\cbIdent}{\defCap{\cbVar_1}{\cbVar_2}}\},\ljStore,(\ljVal\ljAti{1}\conI)\ljAti{2}\con 
      \eval \cbState',\ljStore',\ljBer
    }
    {%
      \ljSEnv \entailsC \cbState,\ljStore, \ljVal\ljAtb{}(\defCap{\conI}{\con}) \eval \cbState',\ljStore',\ljBer
    }\and
    \inferrule [\RuleAssertUnion]
    {%
      \cbVar_1,\cbVar_2\not\in\dom{\cbState}\\
      \ljSEnv \entailsC  \cbState\cup\{\defCb{\cbIdent}{\defCup{\cbVar_1}{\cbVar_2}}\},\ljStore,(\ljVal\ljAti{1}\con)\ljAti{2}\conD 
      \eval \cbState',\ljStore',\ljBer
    }
    {%
      \ljSEnv \entailsC \cbState,\ljStore, \ljVal\ljAtb{}(\defCup{\con}{\conD}) \eval \cbState',\ljStore',\ljBer
    }\and
    \inferrule [\RuleAssertAnd]
    {%
      \cbVar_1,\cbVar_2\not\in\dom{\cbState}\\
      \ljSEnv \entailsC  \cbState\cup\{\defCb{\cbIdent}{\defAnd{\cbVar_1}{\cbVar_2}}\},\ljStore,(\ljVal\ljAti{1}\con)\ljAti{2}\conD 
      \eval \cbState',\ljStore',\ljBer
    }
    {%
      \ljSEnv \entailsC \cbState,\ljStore, \ljVal\ljAtb{}(\defAnd{\con}{\conD}) \eval \cbState',\ljStore',\ljBer
    }\and
    \inferrule [\RuleAssertOr]
    {%
      \cbVar_1,\cbVar_2\not\in\dom{\cbState}\\
      \ljSEnv \entailsC  \cbState\cup\{\defCb{\cbIdent}{\defOr{\cbVar_1}{\cbVar_2}}\},\ljStore,(\ljVal\ljAti{1}\conI)\ljAti{2}\con 
      \eval \cbState',\ljStore',\ljBer
    }
    {%
      \ljSEnv \entailsC \cbState,\ljStore, \ljVal\ljAtb{}(\defOr{\conI}{\con}) \eval \cbState',\ljStore',\ljBer
    }\and
    \inferrule [\RuleAssertNot]
    {%
      \cbVar\not\in\dom{\cbState}\\
      \ljSEnv \entailsC  \cbState\cup\{\defCb{\cbIdent}{\defNeg{\cbVar}}\},\ljStore,\ljVal\ljAti{}\conI 
      \eval \cbState',\ljStore',\ljBer
    }
    {%
      \ljSEnv \entailsC \cbState,\ljStore, \ljVal\ljAtb{}(\defNeg{\conI}) \eval \cbState',\ljStore',\ljBer
    }
  \end{mathpar}
  \caption{Inference rules for contract assertion.}
  \label{fig:lcon_assert}
\end{figure}

It remains to define contract assertion (Figure~\ref{fig:lcon_assert}).

Rule \RefTirName{\RuleLjAssert} applies after the 
first subexpression evaluates to a value and the second subexpression evaluates to a contract-value.
It triggers the evaluation of contract $\con$ in an sandbox environment $\ljSEnv$. 

Asserting a base contract to any value wraps the value to avoid
interference and evaluates the predicate after binding the wrapped value. Finally, a new constraint 
is attached with the result of the predicate.

Once a base contract evaluates to a value (Figure~\ref{fig:lcon_blame}) the constraint state is checked 
whether a violation can be signaled.

Rule $\RefTirName{\RuleAssertDelayed}$ asserts a delayed contract to an object. It wraps its location in a proxy
together with the current sandbox environment, the associated blame identifier, and the contract itself. It is
an error to assert a delayed contract to a primitive value.

Asserting an intersection-contract, an union-contract, or an boolean combination asserts the first subcontract to 
the value before it asserts the second subcontract to the resulting behavior. These contract evaluations create 
a new constraint that contains thecorresponding operator, respectively. 

\begin{figure}[tp]
  \centering
  \begin{mathpar}
    \inferrule [\RuleLjNoBlame]
    {%
      \cbState\textit{ is not a blame state}
    }
    {%
      \ljSEnv \entailsC \cbState,\ljStore,\ljVal\ljAtb{}\ljBer \eval \cbState,\ljStore,\ljVal
    }\and
    \inferrule [\RuleLjBlame]
    {%
      \cbState\textit{ is a blame state}
    }
    {%
      \ljSEnv \entailsC \cbState,\ljStore,\ljVal\ljAtb{}\ljBer \eval \cbState,\ljStore,\ljExn
    }
  \end{mathpar}
  \caption{Inference rules for blame calculation.}
  \label{fig:lcon_blame}
\end{figure}

\subsubsection{Application, Read, and Assignment}

Function application, property read, and property assignment distinguish two cases: either the 
operation applies directly to a non-proxy object or it applies to a proxy. If the target of the 
operation is not a proxy object, then the usual rules apply.

Figure~\ref{fig:lcon_app-get-put} contains the inference rules for function application 
and property access for the non-standard cases.

\begin{figure}[tp]
  \centering
  \begin{mathpar}
    \inferrule [\RuleAppFunction]
    {%
      (\ljLocation', \ljSEnv, \cbIdent, (\defFun{\con}{\conD})) = \ljStore(\ljLocation)\\
      \cbVar_1,\cbVar_2\not\in\dom{\cbState}\\\\
      \ljEnv \entailsFA \cbState\cup\{\defCb{\cbIdent}{\defFun{\cbVar_1}{\cbVar_2}}\},\ljStore,(\ljLocation'(\ljVal\ljAti{1}(\ljSEnv,\con)))\ljAti{2}(\ljSEnv,\conD)
      \eval \cbState',\ljStore',\ljBer
    }
    {%
      \ljEnv \entailsFA \cbState,\ljStore,\ljLocation(\ljVal) \eval 
      \cbState',\ljStore',\ljBer
    }\and
    \inferrule [\RuleAppDependent]
    {%
      (\ljLocation',\ljSEnv,\cbIdent,(\defDep\ljVar\conA)) = \ljStore(\ljLocation)\\
      \ljEnv \entailsFA \cbState,\ljStore,(\ljLocation'(\ljVal))\ljAtb{}\defApp{(\ljSEnv, \conA)}{\ljVal} \eval \cbState',\ljStore',\ljBer
    }
    {%
      \ljEnv \entailsFA \cbState,\ljStore,\ljLocation(\ljVal) \eval \cbState',\ljStore',\ljBer
    }\and
    \inferrule [\RuleAppCap]
    {%
      (\ljLocation', \ljSEnv, \cbIdent, (\defCap{\conQ}{\conR})) = \ljStore(\ljLocation)\\
      \cbVar_1,\cbVar_2\not\in\dom{\cbState}\\\\
      \ljEnv \entailsFA \cbState\cup\{\defCb{\cbIdent}{\defCap{\cbVar_1}{\cbVar_2}}\},\ljStore,((\ljLocation'\ljAti{1}(\ljSEnv,\conQ))\ljAti{2}(\ljSEnv,\conR))\,(\ljVal)
      \eval \cbState',\ljStore',\ljBer
    }
    {%
      \ljEnv \entailsFA \cbState,\ljStore,\ljLocation(\ljVal) \eval 
      \cbState',\ljStore',\ljBer
    }\and
    \inferrule [\RuleAppOr]
    {%
      (\ljLocation', \ljSEnv, \cbIdent, (\defOr{\conQ}{\conR})) = \ljStore(\ljLocation)\\
      \cbVar_1,\cbVar_2\not\in\dom{\cbState}\\\\
      \ljEnv \entailsFA \cbState\cup\{\defCb{\cbIdent}{\defOr{\cbVar_1}{\cbVar_2}}\},\ljStore,((\ljLocation'\ljAti{1}(\ljSEnv,\conQ))\ljAti{2}(\ljSEnv,\conR))\,(\ljVal)
      \eval \cbState',\ljStore',\ljBer
    }
    {%
      \ljEnv \entailsFA \cbState,\ljStore,\ljLocation(\ljVal) \eval 
      \cbState',\ljStore',\ljBer
    }\and
    \inferrule [\RuleAppNeg]
    {%
      (\ljLocation', \ljSEnv, \cbIdent, (\defNeg{\conQ})) = \ljStore(\ljLocation)\\
      \cbVar\not\in\dom{\cbState}\\\\
      \ljEnv \entailsFA \cbState\cup\{\defCb{\cbIdent}{\defNeg{\cbVar}}\},\ljStore,(\ljLocation'\ljAti{}(\ljSEnv,\conQ))\,(\ljVal)
      \eval \cbState',\ljStore',\ljBer
    }
    {%
      \ljEnv \entailsFA \cbState,\ljStore,\ljLocation(\ljVal) \eval 
      \cbState',\ljStore',\ljBer
    }\and
    \inferrule [\RuleGetContract]
    {%
      (\ljLocation',\ljSEnv,\cbIdent,\conM) = \ljStore(\ljLocation)\\
      \ljConst\in\dom{\conM}\\
      \cbVar\not\in\dom{\ljStore}\\\\
      \ljEnv \entailsPR \cbState,\ljStore,(\ljLocation'[\ljConst])\ljAtb{}(\ljSEnv,\conM(\ljConst)) \eval \cbState',\ljStore',\ljBer
    }
    {%
      \ljEnv \entailsPR \cbState,\ljStore,\ljLocation[\ljConst] \eval \cbState',\ljStore',\ljBer
    }\and
    \inferrule [\RulePutContract]
    {%
      (\ljLocation',\ljSEnv,\cbIdent,\conM) = \ljStore(\ljLocation)\\
      \ljConst\in\dom{\conM}\\
      \cbVar\not\in\dom{\ljStore}\\\\
      \ljEnv \entailsPA \cbState\cup\{\defCb{\cbIdent}{\defSet{\cbVar}}\},\ljStore',\ljLocation'[\ljConst]\!=\!(\ljVal\ljAti{}(\ljSEnv,\conM(\ljConst)) \eval \cbState',\ljStore',\ljVal
    }
    {%
      \ljEnv \entailsPA \cbState,\ljStore,\ljLocation[\ljConst]\!=\!\ljVal \eval \cbState',\ljStore',\ljVal
    }
  \end{mathpar}
  \caption{Inference rules for function application, property read, and property assignment.}
  \label{fig:lcon_app-get-put}
\end{figure}

If the target object is a proxy that carries a function contract,
then the domain contract $\C$ is attached to the argument. Next, function application continues 
by passing the input value to the proxy's target location $\ljLocation'$. After completion, the 
range contract $\conD$ is applied to the function's result.  

The assertion of both contracts takes place inside the sandbox environment $\ljSEnv$ stored in the
proxy. A new constraint with two fresh variables indicates that both parts have to hold. 

In case of a dependent contract attached to a function, function
application proceeds by passing the input value to the proxy's target location. Next,
the original function argument gets passed to the contract constructor where the value gets
wrapped to be acceptable in the sandbox environment.
Finally, the wrapped input is bound in the sandbox environment and the contract assertion proceeds
on the function's result.

A property read on a contracted object has two cases depending on the presence of a contract for
the accessed property. If a contract exists, then the contract is attached to the value after
reading it from the target $\ljLocation'$ of the proxy. Otherwise, the value is simply read from
the target. The assertion of a contract happens in the context of the included constraint.

A property write on a contracted object continues with the operation on the target location 
$\ljLocation'$ after the contract is attached to the value. Therefore, the adherence to a 
contract is also checked on assignments, but the check happens in the context of a fresh constraint
that inverts the responsibilities. 

\subsubsection{Sandbox Encapsulation}

\begin{figure}
  \centering
  \begin{mathpar}
    \inferrule [\RuleWrapConst] 
    {}
    {%
      \ljSEnv \entailsWP \cbState,\ljStore,\ljConst \eval \cbState,\ljStore,\ljConst
    }\and
    \inferrule [\RuleWrapContract] 
    {}
    {%
      \ljSEnv \entailsWP \cbState,\ljStore,(\ljSEnv',\con) \eval \cbState,\ljStore,(\ljSEnv',\con)
    }\and
    \inferrule [\RuleWrapConstructor] 
    {}
    {%
      \ljSEnv \entailsWP \cbState,\ljStore,(\ljSEnv',\conA) \eval \cbState,\ljStore,(\ljSEnv',\conA)
    }\and
    \inferrule [\RuleWrapNonProxy]
    {%
      \ljLocation'\not\in\dom{\ljStore}\\
      \not\exists\ljLocation''\in\dom{\ljStore}:~(\ljLocation,\ljSEnv)=\ljStore(\ljLocation'')
    }
    {%
      \ljSEnv \entailsWP \cbState,\ljStore[\ljLocation\mapsto\ljObj],\ljLocation \eval 
      \cbState,\ljStore[\ljLocation\mapsto\ljObj,\ljLocation'\mapsto(\ljLocation,\ljSEnv)],\ljLocation'
    }\and
    \inferrule [\RuleWrapExisting]
    {}
    {%
      \ljSEnv \entailsWP \cbState,\ljStore[\ljLocation'\mapsto(\ljLocation,\ljSEnv)],\ljLocation \eval \cbState,\ljStore,\ljLocation'
    }\and
    \inferrule [\RuleWrapProxy]
    {}
    {%
      \ljSEnv \entailsWP \cbState,\ljStore[\ljLocation'\mapsto(\ljLocation,\ljSEnv)],\ljLocation' \eval \cbState,\ljStore,\ljLocation'
    }\and
    \inferrule [\RuleWrapDelayed]
    {%
      \ljSEnv \entailsWP \cbState,\ljStore[\ljLocation\mapsto(\ljLocation',\ljSEnv,\cbIdent,\conQ)],\ljLocation' \eval
      \cbState,\ljStore',\ljLocation''
    }
    {%
      \ljSEnv \entailsWP \cbState,\ljStore[\ljLocation\mapsto(\ljLocation',\ljSEnv,\cbIdent,\conQ)],\ljLocation \eval
      \cbState,\ljStore'[\ljLocation'''\mapsto(\ljLocation'',\ljSEnv,\cbIdent,\conQ)],\ljLocation''
    }\and
    \inferrule [\RuleAppSandbox] 
    {%
      (\ljLocation',\ljSEnv) = \ljStore(\ljLocation)\\
      (\ljDic, ( \ljEnv', \ljFunction ), \ljVal' ) = \ljStore(\ljLocation')\\\\
      \ljSEnv \entails \cbState,\ljStore,(\ljFunction)(\ljWrap(\ljVal)) \eval \cbState',\ljStore',\ljBer
    }
    {%
      \ljEnv \entailsFA \cbState,\ljStore,\ljLocation(\ljVal) \eval \cbState',\ljStore',\ljBer
    }\and
    \inferrule [\RuleGetSandbox]
    {%
      \ljSEnv \entailsPR \cbState,\ljStore,\ljWrap(\ljLocation'[\ljConst]) \eval \cbState,\ljStore',\ljBer
    }
    {%
      \ljEnv \entailsPR \cbState,\ljStore[\ljLocation\mapsto(\ljLocation',\ljSEnv)],\ljLocation[\ljConst] \eval \cbState,\ljStore'',\ljBer'
    }\and
    \inferrule [\RulePutSandbox]
    {}
    {%
      \ljEnv \entailsPA \cbState,\ljStore[\ljLocation\mapsto(\ljLocation',\ljSEnv)],\ljLocation[\ljConst]\!=\!\ljVal \eval
      \cbState,\ljStore[\ljLocation\mapsto(\ljLocation',\ljSEnv)],\ljExn
    }
  \end{mathpar}
  \caption{Inference rules for sandbox encapsulation.}
  \label{fig:lcon_wrap}
\end{figure}

The sandbox encapsulation (Figure~\ref{fig:lcon_wrap}) distinguishes several cases. A primitive value,
a contract value, and a contract abstraction value is not wrapped. 

To wrap a location that points to a non-proxy object,
the location is packed in a fresh proxy along with the current sandbox environment. 
This packaging ensures that each further access to the wrapped location has to use the current environment.


In case the location is already wrapped by a sandbox proxy or the location of a sandbox proxy gets wrapped 
then the location to the existing proxy is returned. 
This rule ensures that an object is wrapped at most once and 
thus preserves object identity inside the sandbox. 

If the location points to a contracted object, the wrap operation
continues with the target $\ljLocation'$, before adding all contract from the contracted object.


The application of a wrapped function proceeds by unwrapping the function and evaluating it in the
sandbox environment contained in the proxy. The function argument and its result are known to be
wrapped in this case.

A property read on a sandboxed location continues the operation on the target and wraps the
resulting behavior. An assignment to a sandboxed object is not allowed, thus it signals a sandbox violation.